\documentclass[aps,prd,superscriptaddress,amsmath,twocolumn,nofootinbib,10pt,longbibliography]{revtex4-1}
\usepackage[pdftex]{color}
\usepackage[sort&compress]{natbib}
\usepackage[colorlinks=true,linkcolor=blue,filecolor=blue,urlcolor=blue,citecolor=blue,pdftex,plainpages=false]{hyperref}
\usepackage{color,hhline,psfrag,rotating}
\usepackage{dcolumn}
\usepackage{verbatim}
\usepackage{bm}
\usepackage{graphicx}
\usepackage{amsfonts,amssymb,amsmath}
\usepackage{booktabs}  % Improves tables - needed for iPython LaTeX output.
\usepackage{array}
\newcommand{\lamqcd}{\Lambda_\mathrm{QCD}}

\newcommand{\BBsBBdratio}{1.008(25)}
\newcommand{\rzerobags}{0.27(11)}
\newcommand{\rzerobagd}{0.22(12)}
\newcommand{\bhats}{1.232(53)}
\newcommand{\bhatd}{1.222(61)}

\newcommand{\fsqrtbhats}{0.2561(57)\,\mathrm{GeV}}
\newcommand{\fsqrtbhatd}{0.2106(55)\,\mathrm{GeV}}

\newcommand{\Bsmumu}{3.81(18)\times 10^{-9}}
\newcommand{\Bdmumu}{1.031(54)\times 10^{-10}}

\newcommand{\BdmumuBsmumu}{0.02706(70)}
\newcommand{\xifinal}{1.216(16)}
\newcommand{\vtdfinal}{0.00867(23)}
\newcommand{\vtsfinal}{0.04189(93)}
\newcommand{\vratio}{0.2071(27)}

\newcommand{\Gv}{\mathbf{G}}
\newcommand{\Pv}{\mathbf{P}}
\newcommand{\pv}{\mathbf{p}}
\newcommand{\vv}{\mathbf{v}}
\newcommand{\sv}{\mathbf{s}}
\newcommand{\xv}{\mathbf{x}}
\newcommand{\yv}{\mathbf{y}}
\newcommand{\D}{\mathbf{D}}
\newcommand{\cv}{\mathbf{c}} %\boldsymbol{\psi}}
\newcommand{\cov}{\mathbf{M}_\mathrm{cov}}
\newcommand{\corr}{\mathbf{M}_\mathrm{corr}}
\newcommand{\Eq}[1]{Eq.~(\ref{eq:#1})}
\newcommand{\msb}{{\overline{\mathrm{MS}}}}
\newcommand{\nrqcd}{{\mathrm{NR}}}
\newcommand{\latt}{{\mathrm{latt}}}
\newcommand{\nn}{\nonumber}
\newcommand{\be}{\begin{equation}}
\newcommand{\ee}{\end{equation}}
\newcommand{\bea}{\begin{eqnarray}}
\newcommand{\eea}{\end{eqnarray}}

\begin{document}
\title{Neutral B-meson mixing from full  lattice QCD at the physical point }

\author{R.~J.~Dowdall}
\affiliation{DAMTP, University of Cambridge, Wilberforce Road, Cambridge CB3 0WA, UK}
\author{C.~T.~H.~Davies}
\email[]{Christine.Davies@glasgow.ac.uk}
\affiliation{SUPA, School of Physics and Astronomy, University of Glasgow, Glasgow, G12 8QQ, UK}
\author{R.~R.~Horgan}
\affiliation{DAMTP, University of Cambridge, Wilberforce Road, Cambridge CB3 0WA, UK}
\author{G.~P.~Lepage}
\affiliation{Laboratory for Elementary-Particle Physics, Cornell University, Ithaca, NY 14853, USA}
\author{C.~J.~Monahan}
\affiliation{Institute for Nuclear Theory, University of Washington, Seattle, Washington 98195, USA}
\affiliation{Physics Department, College of William and Mary, Williamsburg, Virginia 23187, USA}
\affiliation{Thomas Jefferson National Accelerator Facility, Newport News, Virginia 23606, USA}
\author{J.~Shigemitsu}
\affiliation{Physics Department, The Ohio State University, Columbus, Ohio 43210, USA}
\author{M.~Wingate}
\affiliation{DAMTP, University of Cambridge, Wilberforce Road, Cambridge CB3 0WA, UK}
\collaboration{HPQCD collaboration}
\homepage{http://www.physics.gla.ac.uk/HPQCD}

\date{\today}

\begin{abstract}
We calculate the bag parameters
for neutral $B$-meson mixing in and
beyond the Standard Model, in full four-flavour lattice QCD for the first time.
We work on gluon field configurations that include the effect
of $u$, $d$, $s$ and $c$ sea quarks with the Highly Improved Staggered Quark (HISQ) action
 at three values of the lattice spacing and with three $u/d$ quark masses going down to
the physical value. The valence $b$ quarks use the improved
NRQCD action and the
valence light quarks, the HISQ action.
Our analysis was blinded.
Our results for the bag parameters for all five operators are the
most accurate to date. For the Standard Model operator between
$B_s$ and $B_d$ mesons we find: $\hat{B}_{B_s}=\bhats$, $\hat{B}_{B_d}=\bhatd$.
Combining our results with lattice QCD calculations of the decay
constants using HISQ quarks from the Fermilab/MILC collaboration and with
experimental values for $B_s$ and $B_d$ oscillation
frequencies allows determination of the CKM elements $V_{ts}$ and $V_{td}$.
We find $|V_{ts}| = \vtsfinal$, $|V_{td}| = \vtdfinal$ and $|V_{ts}|/|V_{td}| = \vratio$.
Our results agree well (within $2\sigma$) with values determined from
CKM unitarity constraints based on tree-level processes (only). Using a ratio
to $\Delta M_{s,d}$ in which CKM elements cancel in the Standard Model,
we determine the branching fractions
${\text{Br}}(B_s\rightarrow \mu^+\mu^-) = \Bsmumu$ and
${\text{Br}}(B_d\rightarrow \mu^+\mu^-) = \Bdmumu$.
We also give
results for matrix elements of the operators $R_0$, $R_1$
and $\tilde{R}_1$ that contribute to
neutral $B$-meson width differences.
\end{abstract}

%\pacs{}
%\keywords{}

\maketitle

%%%%%%%%%%%%%%%%%%%%%%%%%%%%%%%%%%%%%%%%%%%%%%%%%%%%%%%%%%%%%%%%%
%
\section{Introduction}
\label{sec:intro}
%
%%%%%%%%%%%%%%%%%%%%%%%%%%%%%%%%%%%%%%%%%%%%%%%%%%%%%%%%%%%%%%%%%+
The Standard Model description of neutral $B_d$ and $B_s$ oscillations
requires knowledge of hadronic parameters derived from the matrix
elements of 4-quark operators between $B_q$ and $\overline{B}_q$
states. These 4-quark operators come from the effective electroweak Lagrangian
at energy scales appropriate to $B$ physics and the matrix elements
can only be determined by lattice
QCD calculations,
which are now able to include the full impact of QCD
on such hadronic quantities~\cite{Davies:2003ik}.
The accuracy with which this can be done
is the limiting factor in the constraint that can be obtained from
the now very precise experimental results on the neutral meson mass
difference (seen as an oscillation frequency). In the
Standard Model this constraint leads to a determination of
the Cabibbo-Kobayashi-Maskawa (CKM)
matrix elements that accompany the 4-quark operators of the
Standard Model.
New physics models with extra heavy particles
extend the effective Hamiltonian to include additional
4-quark operators. Constraints on the new physics from experiment then
need accurate determination of the matrix elements of the
new operators. Again this can come only from lattice QCD calculations.

Here we provide the first ``second-generation'' lattice QCD calculation of
the matrix elements of all five $\Delta B$ = 2 operators
of dimension six for the $B_s$ and $B_d$.
We improve on earlier calculations by working on
gluon field configurations generated by the MILC collaboration
that include $u$, $d$, $s$
and $c$ quarks in the sea with $u/d$ quark
masses going down to their physical values.
Although this obviates the need for a chiral
extrapolation, we also include heavier $u/d$ quark masses in our
set of results so that we can map out the dependence on
the light quark mass. The discretisation of QCD that we
use is
fully improved through $\mathcal{O}(\alpha_sa^2)$
for both the gluons and the light quarks (including all
of those in the sea) for which the Highly Improved
Staggered Quark (HISQ) action is used. For the $b$ quarks we use
improved NonRelativistic QCD, which includes $\mathcal{O}(\alpha_s)$
corrections to terms at order $v^4$ (where $v$ is the heavy
quark velocity). By linking this calculation directly to
our earlier one for the $B$-meson decay constants~\cite{Dowdall:2013tga} (that
parameterize the amplitude to create a meson from the vacuum)
we are able to give results directly for the ``bag parameters''
associated with each operator and take advantage of the cancellation of
a number of systematic effects. The bag parameters
(to be defined in Section~\ref{sec:cont4q}) encode the
multiplicative factor by which the operator matrix element
differs from that expected in the vacuum saturation approximation,
which is related to the decay constant.
To the extent to which this approximation works (and we will
show here that it does work well) we expect the bag parameters
to have very little dependence on light quark sea or valence
masses and even on the lattice spacing. This enables improvements
in accuracy over earlier work along with a much simpler picture
of the extrapolation to the physical point.

The first unquenched lattice QCD calculations of the matrix elements
for neutral $B$ meson mixing focussed purely on results
for the Standard Model
operators~\cite{Dalgic:2006gp, Gamiz:2009ku} and ratios
for $B_s$ to $B_d$~\cite{Bazavov:2012zs}. Calculations have also
been done in the infinite heavy quark mass limit~\cite{Aoki:2014nga}.
More recently calculations of matrix elements for the full set
of SM and BSM operators have been
done~\cite{Carrasco:2013zta, Bazavov:2016nty}.
The calculations in~\cite{Carrasco:2013zta} use the twisted mass formalism
for all quarks on gluon field configurations including $u$ and $d$
quarks in the sea. An extrapolation of results (renormalised using the
RI-MOM scheme)
is made from a heavy quark mass in the charm region
up to the $b$ quark mass using ratios with a known infinite
mass limit.
The calculations in~\cite{Bazavov:2016nty} use the Fermilab
formalism for the $b$ quark and the asqtad formalism for the
light quarks on gluon field configurations that include
$u$, $d$ and $s$ quarks in the sea with the asqtad formalism.
Perturbatively renormalised 4-quark operator matrix elements (only)
are calculated and so bag parameters
must be derived using decay constant results from elsewhere.
A very recent result in~\cite{Boyle:2018knm} uses domain-wall
quarks on gluon field configurations including $u$, $d$ and $s$
in the sea and extrapolates in heavy quark mass
to the $b$ quark mass for $B_s$ to $B_d$ ratios for SM mixing
matrix elements and decay constants.

In Section~\ref{sec:background} we discuss the 4-quark operators relevant
to $B$~mixing and how they are implemented on the lattice.
Section~\ref{sec:lattice} describes our lattice calculation and
results. We compare our results to previous work
in Section~\ref{sec:discussion}, determine CKM elements $V_{ts}$
and $V_{td}$ using experimental results on $B$-meson
mass differences, determine branching fractions for the rare decays
of $B_d$ and $B_s$ to $\mu^+\mu^-$, and give matrix elements for derived operators
that contribute to width differences. Section~\ref{sec:conclusions}
gives our conclusions and discusses the prospects for future
improvements. Details about our analysis are
contained in four Appendices:
Appendix~\ref{sec:fitting-protocols} on the
lattice QCD correlators that we calculate and how we fit them
to obtain matrix elements and bag parameters;
Appendix~\ref{sec:chifits} on
the chiral perturbation theory fits we use to combine results
at physical and unphysical light quark masses;
Appendix~\ref{sec:corr} on correlations in the uncertainties for our
final results
and lastly, with more general applications beyond this analysis,
Appendix~\ref{sec:SVD-cuts}
on SVD cuts and fitting correlators.

%%%%%%%%%%%%%%%%%%%%%%%%%%%%%%%%%%%%%%%%%%%%%%%%%%%%%%%%%%%%%%%%
%
\section{Background}
\label{sec:background}

\subsection{Continuum 4-quark operators}
\label{sec:cont4q}

\begin{figure}
\includegraphics[width=0.95\hsize]{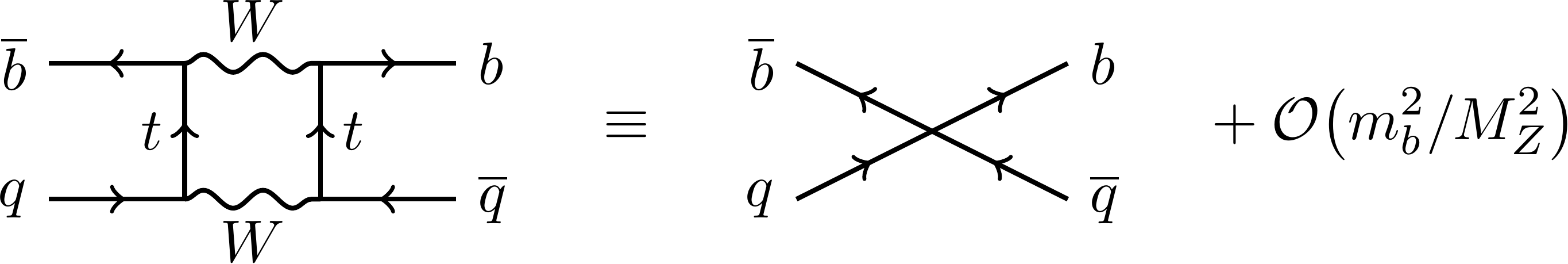}
\caption{\label{fig:box-pic}
An example from the Standard Model of a mechanism that
mixes the neutral $B_q$ and~$\overline{B}_q$. The amplitude
is well approximated by a contact term for matrix elements
between $B_q$-meson states.}
\end{figure}

Neutral $B$-meson mixing occurs at lowest order in the Standard Model
through box diagrams involving the exchange of $W$ bosons and top
quarks, see Figure~\ref{fig:box-pic}.
These box diagrams can be well approximated by an effective
Lagrangian expressed in terms of 4-quark operators. Here we will
examine all five of the independent local dimension-6 operators
that could contribute to $\Delta B=2$
processes~\cite{Gabbiani:1996hi,Bazavov:2016nty}:
\begin{eqnarray}
\label{eq:Odef}
O_1 &=& \left[ \overline{\Psi}_b^i(V-A)\Psi_q^i\right] \left[\overline{\Psi}_b^j(V-A)\Psi_q^j\right] \nonumber \\
O_2 &=& \left[ \overline{\Psi}_b^i(S-P)\Psi_q^i\right] \left[\overline{\Psi}_b^j(S-P)\Psi_q^j\right] \nonumber \\
O_3 &=& \left[ \overline{\Psi}_b^i(S-P)\Psi_q^j\right] \left[\overline{\Psi}_b^j(S-P)\Psi_q^i\right] \nonumber \\
O_4 &=& \left[ \overline{\Psi}_b^i(S-P)\Psi_q^i\right] \left[\overline{\Psi}_b^j(S+P)\Psi_q^j\right] \nonumber \\
O_5 &=& \left[ \overline{\Psi}_b^i(S-P)\Psi_q^j\right] \left[\overline{\Psi}_b^j(S+P)\Psi_q^i\right]
\end{eqnarray}
where $V=\gamma_{\mu}$, $A=\gamma_{\mu}\gamma_5$, $S=1$ and $P=\gamma_5$,
and sums over~$\mu$ and color indices~$i$ and~$j$ are implicit.
In the Standard Model, the most important of
these for $B$-$\overline{B}$~mixing is~$O_1$. This operator mixes with
$O_2$ under renormalization. Operators~$O_4$ and~$O_5$ do not
appear in the Standard Model, but do arise in various BSM scenarios.

It is conventional to parameterize matrix elements of these operators
in terms of ``bag parameters,''
\begin{equation}\label{eq:bagparam}
    B_{B_q}^{(i)}(\mu) \equiv
    \frac{\langle {B}_q | O_i^q | \overline{B}_q\rangle_\msb^{(\mu)}}{\eta^q_i(\mu) f_{B_q}^2 M_{B_q}^2},
\end{equation}
where here $\mu$ is the renormalization scale,
and $M_{B_q}$ and $f_{B_q}$ are the mass and weak decay constant of
the $B_q$~meson:
\begin{equation}
\label{eq:fdef}
\langle 0 | \overline{\Psi}_q^i\gamma_0\gamma_5\Psi_b^i | B_q(\vec{p}=0)\rangle = f_{B_q}M_{B_q}.
\end{equation}
The normalization parameter~$\eta_i^q(\mu)$ is chosen so that the bag parameters
equal~1 in the ``vacuum saturation approximation,'' where gluon (and other
QCD) exchanges
between the initial and final $\overline{B}_q$ and ${B}_q$ are
ignored (see~\cite{Gabbiani:1996hi,Bazavov:2016nty} for more details):
\begin{eqnarray}
\label{eq:etadef}
\eta_1^q  &=&  \frac{8}{3}  \\
\eta_2^q  &=&  -\frac{5}{3} \left(\frac{M_{B_q}}{m_b(\mu)+m_q(\mu)}\right)^2 \nonumber \\
\eta_3^q  &=&  \frac{1}{3} \left(\frac{M_{B_q}}{m_b(\mu)+m_q(\mu)}\right)^2 \nonumber \\
\eta_4^q  &=&  2 \left[\left(\frac{M_{B_q}}{m_b(\mu)+m_q(\mu)}\right)^2+\frac{1}{6}\right] \nonumber \\
\eta_5^q  &=&  \frac{2}{3} \left[\left(\frac{M_{B_q}}{m_b(\mu)+m_q(\mu)}\right)^2+\frac{3}{2}\right] . \nonumber
\end{eqnarray}
We use renormalization scale~$\mu=m_b(m_b)$; the corresponding values
for the normalization factors are given in Table~\ref{tab:eta_iq}.

\begin{table}
\caption{\label{tab:eta_iq} Normalizations $\eta_i^q(m_b)$ for bag
    parameters (\Eq{etadef}). These are calculated using
    $M_{B_s}$=5.3669(2)\,GeV
    and $M_{B_d}$=5.2796(2)\,GeV~\cite{Tanabashi:2018oca},
    $\overline{m}_b(\overline{m}_b)$ = 4.162(48)\,GeV
    and $m_b/m_s$=52.55(55)~\cite{Chakraborty:2014aca}, and
    $m_s/m_l$=27.18(10)~\cite{Bazavov:2017lyh}.}
\begin{ruledtabular}
\begin{tabular}{cccccc}
$B_q$ & $\eta_1^q$ & $\eta_2^q$ & $\eta_3^q$ & $\eta_4^q$ & $\eta_5^q$ \\
\hline \hline
$B_s$ & 2.667 & $-2.669\,(62)$ & 0.534\,(12) & 3.536\,(74) & 2.068\,(25) \\
$B_d$ & 2.667 & $-2.678\,(62)$ & 0.536\,(12) & 3.547\,(74) & 2.071\,(25) \\
\end{tabular}
\end{ruledtabular}
\end{table}

The bag parameters provide both computational advantages and physical insights.
The leading-order logarithms in chiral perturbation theory, coming from
the matrix element of the 4-quark operator and~$f_{B_q}^2$, partly cancel in the ratio;
see Appendix~\ref{sec:chifits}. In particular the coefficient of the chiral
logarithm from the
tadpole diagrams is reduced by a factor of 4. Therefore
bag parameters should be less dependent upon the light-quark mass;
we find very little
mass dependence. Finite-volume effects will be correspondingly
reduced. We also find that most of the dependence on lattice spacing
cancels. Finally, as we will show, the bag parameters all turn out to be
of order one,
suggesting that vacuum saturation is a useful approximation.
For these reasons, we focus here on bag parameters; values for the
matrix elements are easily obtained from the bag parameters
given values for the decay constants~\cite{Dowdall:2013tga,Bazavov:2017lyh}.

%%%%%%%%%%%%%%%%%%%%%%%%%%%%%%%%%%%%%%%%%%%%%%%%%%%%%%%%%%%%%%%%%
\subsection{Lattice QCD 4-quark operators and matching}
\label{sec:lat4q}

Matrix elements of the 4-quark operators are regulator dependent, and
so we need to convert matrix elements calculated in our simulation (with the
lattice regulator) into the corresponding matrix elements for the more
conventional $\msb$~scheme. The differences between the two schemes are
ultraviolet and so can be calculated using QCD perturbation theory.
To lowest and first order in $\alpha_s$ the relationship has the form
(for $\mu=m_b$):
\begin{align}\label{eq:matching}
    \langle O_i \rangle_\msb^{(m_b)}
    &= \big(1 + \alpha_s\, z_{ii}\big) \langle O_i \rangle_\mathrm{latt} \nn\\
     &\quad+ \sum_{j\ne i} \alpha_s\, z_{ij} \langle O_j \rangle_\mathrm{latt}\nn\\
     &\quad+\mathcal{O}\Big(\alpha_s^2, \frac{\alpha_s\Lambda_\mathrm{QCD}}{m_b}, \alpha_s (a\Lambda_\mathrm{QCD})^2\Big) \, .
\end{align}
The coefficients~$z_{ij}$ relevant to our simulation were calculated
in~\cite{Monahan:2014xra} and are summarized in Table~\ref{tab:zresults}.
The scale for $\alpha_s$ depends on the lattice spacing; we use the same
values for~$\alpha_s$
used in~\cite{Colquhoun:2015oha} to calculate renormalizations
for the axial-vector current that couples to $B_q$~mesons
(see Table~\ref{tab:params} for the values).

\begin{table*}
\caption[jgkj]{\label{tab:zresults}
Perturbative coefficients used in \Eq{matching} to convert
matrix elements of lattice NRQCD-HISQ 4-quark operators into
$\msb$~matrix elements.
Results are given for the NRQCD valence
$b$-quark masses (in lattice units) used with our different ensembles.
The continuum scheme used is the $\msb_\mathrm{NDR}$ scheme
of~\cite{Beneke:1998sy} (BBGLN) with  $\mu=m_b$.
The coefficients come from~\cite{Monahan:2014xra}, with
$z_{ij}\equiv\rho_{ij}-\zeta_{ij}$ where $\rho_{ij}$ and $\zeta_{ij}$
are listed in Tables~III and~IV of that paper\footnote{Note that we
have corrected two typographical errors, for $\rho_{21}$ for $am_b=2.66$
and $\zeta_{22}$ for $am_b=2.62$.}.  The perturbative
coefficients~$z_{A_0}$ for the temporal axial current (\Eq{A0matching})
are also listed; these are from~\cite{Dowdall:2013tga}, which used
results from~\cite{Monahan:2012dq}.
}

\begin{ruledtabular}
\begin{tabular}{cccccccccccc}
$am_b$ & $z_{11}$ & $z_{12}$ & $z_{22}$ & $z_{21}$ & $z_{33}$ & $z_{31}$ & $z_{44}$ & $z_{45}$ & $z_{55}$ & $z_{54}$ & $z_{A_0}$ \\
\hline \hline
3.297 & $-0.472\,(2)$ & $-0.299\,(2)$ & 0.440\,(2) & 0.041\,(2) & 0.036\,(2) & 0.092\,(2) & 0.646\,(2) & $-0.252\,(2)$ & $-0.141\,(2)$ & $0.111\,(2)$ & $0.024\,(2)$ \\
3.263 & $-0.469\,(2)$ & $-0.296\,(2)$ & 0.438\,(2) & 0.041\,(2) & 0.038\,(2) & 0.091\,(2) & 0.640\,(2) & $-0.251\,(2)$ & $-0.140\,(2)$ & $0.108\,(2)$ & $0.022\,(2)$ \\
3.25  & $-0.469\,(2)$ & $-0.294\,(2)$ & 0.438\,(2) & 0.041\,(2) & 0.040\,(2) & 0.091\,(2) & 0.639\,(2) & $-0.252\,(2)$ & $-0.139\,(2)$ & $0.106\,(2)$ & $0.022\,(2)$ \\
2.66  & $-0.429\,(2)$ & $-0.235\,(2)$ & 0.394\,(2) & 0.044\,(2) & 0.101\,(2) & 0.080\,(2) & 0.514\,(2) & $-0.254\,(2)$ & $-0.127\,(2)$ & $0.037\,(2)$ & $0.006\,(2)$ \\
2.62  & $-0.427\,(2)$ & $-0.229\,(2)$ & 0.388\,(2) & 0.044\,(2) & 0.105\,(2) & 0.080\,(2) & 0.501\,(2) & $-0.254\,(2)$ & $-0.128\,(2)$ & $0.032\,(2)$ & $0.001\,(2)$ \\
1.91  & $-0.296\,(2)$ & $-0.108\,(2)$ & 0.340\,(2) & 0.045\,(2) & 0.259\,(2) & 0.053\,(2) & 0.299\,(2) & $-0.243\,(2)$ & $-0.063\,(2)$ & $-0.084\,(2)$ & $-0.007\,(2)$ \\
\end{tabular}
\end{ruledtabular}

\end{table*}

Our lattice analysis uses non-relativistic QCD (NRQCD) for the $b$~dynamics.
Quarks and anti-quarks decouple in NRQCD, and so correspond to separate
fields. As a result the lattice version of a 4-quark operator has the
form~\cite{Monahan:2014xra}
\begin{align}
[\overline{\Psi}_b\Gamma_1\Psi_q][\overline{\Psi}_b\Gamma_2&\Psi_q] \to %\nn\\
    [\overline{\Psi}_b^\nrqcd\Gamma_1\Psi_q][\overline{\Psi}_{\overline{b}}^\nrqcd\Gamma_2\Psi_q]
    \nn \\
    &+ \frac{1}{2m_b}\big[\bm{\nabla}\overline{\Psi}_{b}^\nrqcd\cdot\bm{\gamma}\Gamma_1\Psi_q\big]
                     \big[\overline{\Psi}_{\overline{b}}^\nrqcd\Gamma_2\Psi_q\big]
    \nn \\
    &+ \frac{1}{2m_b}\big[\overline{\Psi}_{b}^\nrqcd\Gamma_1\Psi_q\big]
                     \big[\bm{\nabla}\overline{\Psi}_{\overline{b}}^\nrqcd\cdot\bm{\gamma}\Gamma_2\Psi_q\big]
    \nn \\
    &+ \big(\Gamma_1 \leftrightarrow \Gamma_2\big),
\end{align}
where $\overline{\Psi}_b^\nrqcd$ creates a $b$~quark and
$\overline{\Psi}_{\overline{b}}^\nrqcd$ annihilates a $b$~anti-quark.
These are the lattice operators we use on the right-hand side of \Eq{matching}.
The $1/m_b$ terms are the $\mathcal{O}(\lamqcd/m_b)$ corrections
to the operator in NRQCD.

\begin{figure}
\includegraphics[width=0.8\hsize]{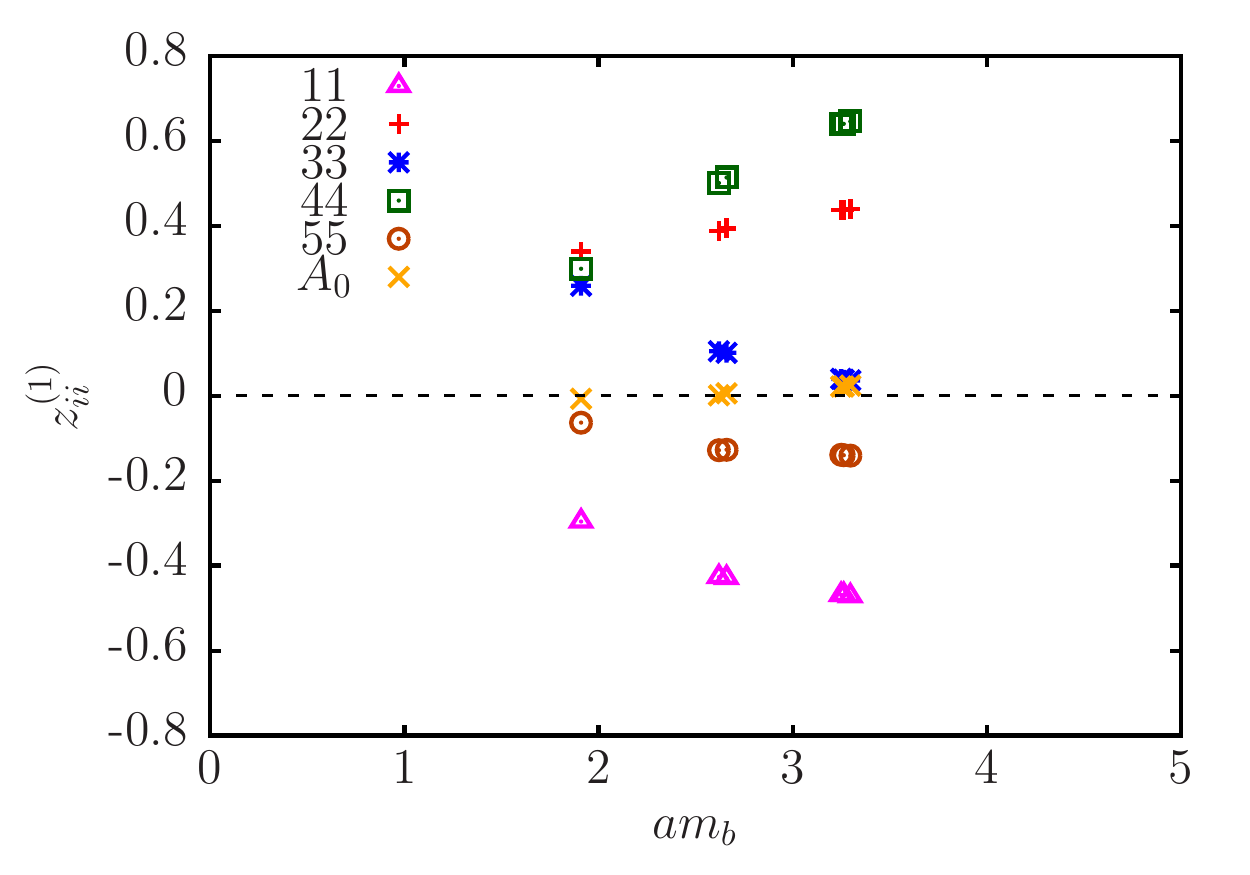}
\includegraphics[width=0.8\hsize]{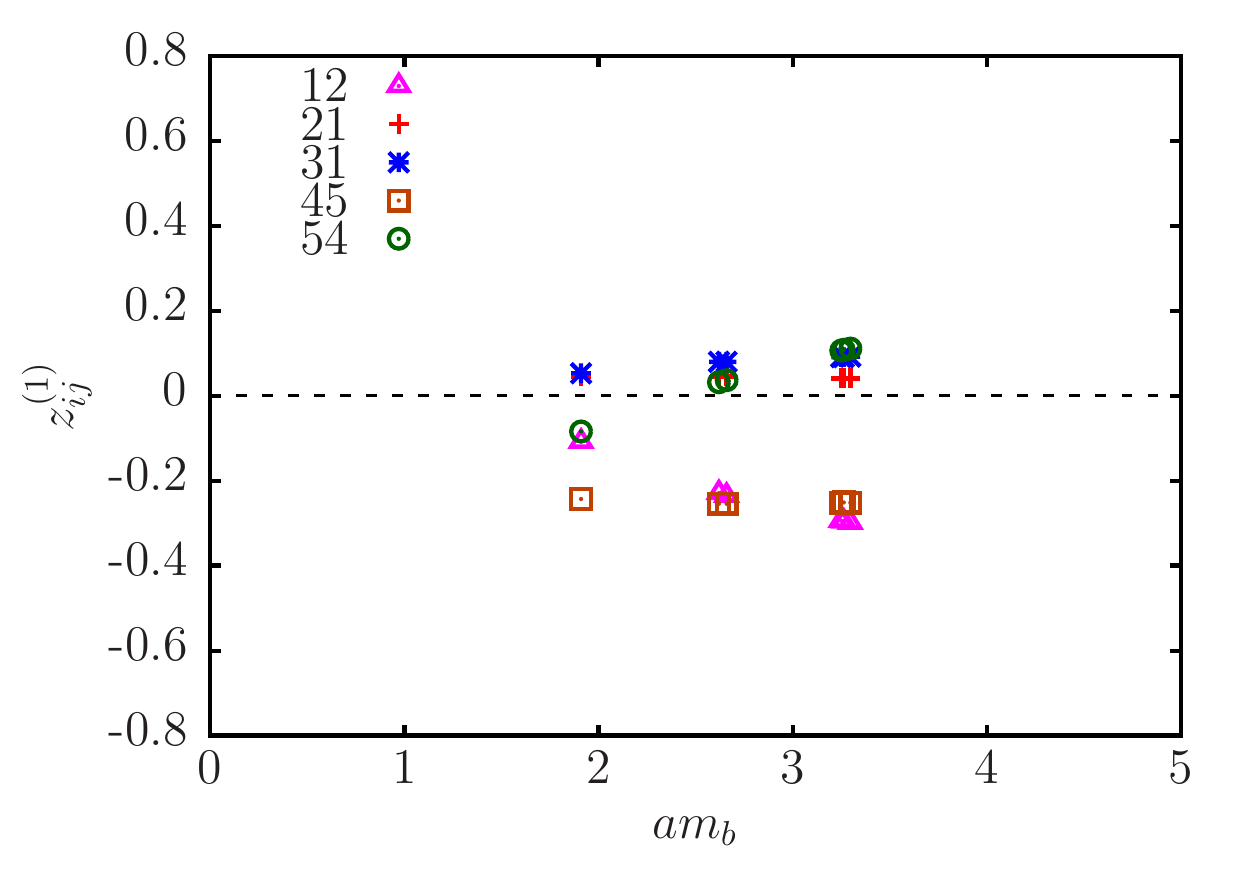}
\caption{\label{fig:zplot} Coefficients~$z_{ij}$ of $\mathcal{O}(\alpha_s)$
terms in the matching of lattice NRQCD-HISQ 4-quark operators
to the $\msb$ scheme plotted as a function of the bare
NRQCD $b$ quark mass in lattice units. The top plot shows the
diagonal coefficients ($i=j$) that enter the renormalization of
a given operator; the lower
plot shows the off-diagonal coefficients ($i\ne j$) corresponding
to the mixing of different operators.
See \Eq{matching} for the definition of $z_{ij}$ and Table~\ref{tab:zresults}
for the values. The $ij$~values are
indicated in the key. ``$A_0$'' refers to the $\mathcal{O}(\alpha_s)$
coefficient for the renormalization of the temporal axial current ($z_{A_0}$
in Table~\ref{tab:zresults}). }
\end{figure}

A complication for operators $O_2$ and $O_3$ is the treatment of
``evanescent operators.'' Our matching results use the $\msb_\mathrm{NDR}$
scheme
of~\cite{Beneke:1998sy}~(BBGLN). These matrix elements are readily
converted
to the alternative scheme of~\cite{Buras:2001ra} (BJU) using
the following equations  (through $\mathcal{O}(\alpha_s)$, with
$\mu=m_b$)~\cite{Becirevic:2001xt, Ciuchini:2003ww}:
\begin{align}
\label{eq:convert}
{{O}}_2\Big|_{\mathrm{BJU}} &= {{O}}_2 +
 \alpha_s \left( -0.318 \,{{O}}_2 -0.013 \,{{O}}_1\right)\Big|_{\mathrm{BBGLN}} \nonumber \\
{{O}}_3\Big|_{\mathrm{BJU}} &= {{O}}_3 +
 \alpha_s \left( 0.106 \,{{O}}_3 -0.013 \,{{O}}_1\right)\Big|_{\mathrm{BBGLN}} \, .
\end{align}

The matching coefficients~$z_{ij}$ from \Eq{matching} are plotted against
$am_b$ in Figure~\ref{fig:zplot}. These coefficients are
not large and have a relatively benign dependence
on the $b$-quark mass across the range that we use, although different
coefficients behave differently.
The diagonal coefficients~$z_{ii}$
are much larger than the corresponding coefficients for the
NRQCD-HISQ axial current ($z_{A_0}$ in Table~\ref{tab:zresults}),
which are unusually small.
Note that the only nonzero
off-diagonal coefficients are for $ij$ equal
to 12, 21, 31, 45, and~54, and that these tend to be smaller than the diagonal
parameters.

It is worth remarking here on the similarities and differences between the perturbative
matching we apply here and that used by the Fermilab/MILC
collaborations~\cite{Bazavov:2016nty}
in their determination of $B$ mixing matrix elements. They also make use of a
perturbative calculation of the matching to $\mathcal{O}(\alpha_s)$. They do this
after a non-perturbative determination of factors that are needed to remove normalisation
artifacts from the clover and asqtad actions that they use for heavy and light
quarks respectively and without which they would have large $\mathcal{O}(\alpha_s)$
coefficients. We do not need to apply this procedure because the NRQCD and
HISQ actions are well-behaved in this respect~\cite{Chakraborty:2017hry}.
After applying their nonperturbative procedure, the Fermilab/MILC collaboration
give results for their $\mathcal{O}(\alpha_s)$ coefficients in Table III of~\cite{Bazavov:2016nty}.
Their coefficients differ from ours because they are using a different discretization
of QCD for both heavy and light quarks. However qualitatively the coefficients show
similar behavior in terms of magnitude and dependence on the lattice $b$ quark
mass (given in their case by the parameter $\kappa_b^{\prime}$).

%%%%%%%%%%%%%%%%%%%%%%%%%%%%%%%%%%%%%%%%%%%%%%%%%%%%%%%%%%%%%%%%
%
\section{Lattice Calculation}
\label{sec:lattice}
%
%%%%%%%%%%%%%%%%%%%%%%%%%%%%%%%%%%%%%%%%%%%%%%%%%%%%%%%%%%%%%%%%%
\subsection{Simulations}
\begin{table}
\caption{
Parameters of the gauge ensembles used in this calculation.
$\beta$ is the gauge coupling, $a_{\Upsilon}$ is the lattice spacing as
determined by the $\Upsilon(2S-1S)$ splitting in \cite{Dowdall:2011wh},
where the three errors are statistics, NRQCD systematics and experiment.
$am_l,am_s$ and $am_c$ are the sea quark masses, $L \times T$ gives the spatial and temporal extent of the lattices and $n_{{\rm cfg}}$ is the number of configurations in each ensemble.
 We use 16 time sources on each configuration to improve statistics.
}
\label{tab:gaugeparams}
\begin{ruledtabular}
\begin{tabular}{llllllllll}
Set & $\beta$ & $a_{\Upsilon}$ (fm) 	& $am_{l}$ & $am_{s}$ & $am_c$ & $L \times T$ & $n_{{\rm cfg}}$  \\
\hline\hline
1 & 5.8 & 0.1474(5)(14)(2)  & 0.013   & 0.065  & 0.838 & 16$\times$48 & 1000 \\
2 & 5.8 & 0.1463(3)(14)(2)  & 0.0064  & 0.064  & 0.828 & 24$\times$48 & 1000 \\
3 & 5.8 & 0.1450(3)(14)(2)  & 0.00235 & 0.0647 & 0.831 & 32$\times$48 & 1000
\\[1ex]
4 & 6.0 & 0.1219(2)(9)(2)   & 0.0102  & 0.0509 & 0.635 & 24$\times$64 & 1000 \\
5 & 6.0 & 0.1195(3)(9)(2)   & 0.00507 & 0.0507 & 0.628 & 32$\times$64 & 1000 \\
6 & 6.0 & 0.1189(2)(9)(2)   & 0.00184 & 0.0507 & 0.628 & 48$\times$64 & 1000
\\[1ex]
7 & 6.3 & 0.0884(3)(5)(1)   & 0.0074  & 0.037  & 0.440 & 32$\times$96 & 1007 \\
\end{tabular}
\end{ruledtabular}
\end{table}

\begin{table}
\caption{
\label{tab:params}
Parameters used for the valence quarks. $am_b$ is the bare $b$ quark mass in lattice units, $u_{0L}$ is the Landau link value used for tadpole-improvement,
and $am_l^{\rm val}$, $am_s^{\rm val}$ are the HISQ light and strange quark valence masses.
We also tabulate the values from~\cite{Colquhoun:2015oha} for the running
coupling constant~$\alpha_s$ to be used in \Eq{matching} for matching
lattice 4-quark operators to the continuum. This is in the V-scheme
at scale $(2/a)$.
}
\begin{ruledtabular}
\begin{tabular}{llllll}
Set & $am_b$ & $u_{0L}$ & $am_l^{\rm val}$ & $am_s^{\rm val}$ & $\alpha_s$ \\
\hline\hline
1 & 3.297 & 0.8195   & 0.013   & 0.0641 & 0.346\\
2 & 3.263 & 0.82015  & 0.0064  & 0.0636 & 0.345\\
3 & 3.25  & 0.819467 & 0.00235 & 0.0628 & 0.343
\\[1ex]
4 & 2.66 & 0.834    & 0.01044 & 0.0522  & 0.311\\
5 & 2.62 & 0.8349   & 0.00507 & 0.0505  & 0.308\\
6 & 2.62 & 0.834083 & 0.00184 & 0.0507  & 0.307
\\[1ex]
7 & 1.91 & 0.8525   & 0.0074  & 0.0364  & 0.267
\end{tabular}
\end{ruledtabular}
\end{table}

We use seven ensembles of gluon field configurations
recently generated by the MILC collaboration~\cite{Bazavov:2010ru,Bazavov:2012xda}.
Details are given in Table~\ref{tab:gaugeparams}.
We use ensembles at three values of the lattice spacing, $a$, to control
discretisation effects and at three values of the light quark mass down
to the physical point to map out sea quark mass effects. 
Discretisation effects depend on $a^2$ and sea quark mass effects 
are approximately linear, so a range in $a^2$ of 
a factor of 3 and in 
sea light quark mass of a factor of 5 allows us substantial leverage to 
pin down these effects.  
The lattice spacing values were
determined using the mass splitting between the $\Upsilon$ and $\Upsilon^{\prime}$,
as described in~\cite{Dowdall:2011wh} where a discussion of systematic errors
can be found.
The sea quarks use HPQCD's HISQ action~\cite{Follana:2006rc} which we have
shown to have small discretisation errors
even for charm quarks~\cite{oldfds, newfds, Donald:2012ga}. This enables
four flavors of quarks to be included in the sea, with masses given
in Table~\ref{tab:gaugeparams}. The $u$ and $d$ quark masses
are taken to be the same.

The valence $b$ quarks are implemented using lattice
NonRelativistic QCD (NRQCD)~\cite{Lepage:1992tx}.
The action is described in detail in~\cite{Dowdall:2011wh}.
It includes a number of improvements over earlier calculations,
in particular one-loop radiative corrections
(beyond tadpole-improvement~\cite{Lepage:1992xa})
to most of
the coefficients of the $\mathcal{O}(v_b^4)$ relativistic correction terms.
The tadpole-improvement of the action is done using the Landau gauge-link,
with $u_{0L}$ values given in Table~\ref{tab:params}.
This action has been shown to give excellent agreement with experiment in
recent calculations of the bottomonium~\cite{Dowdall:2011wh,Daldrop:2011aa, Dowdall:2013jqa}
and $B$-meson spectra~\cite{Dowdall:2012ab}.
The $b$ quark mass is tuned, giving the values in Table~\ref{tab:params}, by
fixing the spin-averaged kinetic mass of the $\Upsilon$ and $\eta_b$ states
to experiment~\cite{Dowdall:2011wh}. 
NRQCD breaks down as $am_b \rightarrow 0$ 
but all our values of $am_b$ are substantially larger than 1, 
where there is no problem.

The HISQ valence light quark masses are taken to be equal
to the sea mass except on set 4 where there is a slight discrepancy.
The $s$ quark is tuned using the mass of the $\eta_s$ meson~\cite{Davies:2009tsa},
a fictitious pseudoscalar $s\overline{s}$ state which is not
allowed to decay on the lattice. Its properties can be
very accurately determined in lattice QCD and we find
$M_{\eta_s} =$ 0.6885(22) GeV~\cite{Dowdall:2013rya}.
Values for valence $s$ masses are given in Table~\ref{tab:params} and
corresponding values of $M_{\eta_s}$ in lattice units
in~\cite{Dowdall:2013rya}. We allow for uncertainties from
mistuned valence masses in our determination of the physical results.

\subsection{Simulation Results and Error Budget}
\label{sec:simerr}
We describe the 2-point and 3-point correlators used in our analysis
in Appendix~\ref{sec:fitting-protocols}. We use the 2-point correlators
to extract the decay constants~$f_{B_q}$, including
the $1/m_b$~corrections (\Eq{A0matching}). We also combine them with the
3-point correlators to calculate lattice matrix elements of the~$O_n$.
Also in that Appendix, we discuss the
Bayesian fits used to extract physics from these correlators.
Our final results
for $\langle {B}_q | O_n | \overline{B}_q \rangle_\latt /(f_{B_q} M_{B_q})^2$
are summarized in Table~\ref{tab:fit-results} of
Appendix~\ref{sec:fitting-protocols}.
As discussed in the Appendix, this was a blind analysis.

\begin{table}
\caption{\label{tab:OnMSB} $\msb$ matrix elements obtained
from \Eq{matching} together with
simulation data from Table~\ref{tab:fit-results}. Results are given
for each configuration set and both $B_s$ (top) and $B_d$~(bottom) mesons.
Values are also given (``phys.'') for our final results at
physical quark masses.
}

\begin{ruledtabular}
\begin{tabular}{cccccc}
\\[-2ex]
 & \multicolumn{5}{c}{$\langle B_s | O_n | \overline{B}_s\rangle_\msb^{(m_b)}/(f_{B_s}M_{B_s})^2$} \\[1ex]
set & $O_1$ & $O_2$ & $O_3$ & $O_4$ & $O_5$ \\
\hline\hline
1 & 2.10\,(21) & $-2.14\,(22)$ & 0.442\,(59) & 3.56\,(41) & 1.90\,(14)   \\
2 & 2.16\,(21) & $-2.20\,(22)$ & 0.441\,(59) & 3.73\,(43) & 2.00\,(14)   \\
3 & 2.14\,(21) & $-2.18\,(22)$ & 0.441\,(58) & 3.69\,(41) & 1.97\,(14)  \\[1ex]
4 & 2.20\,(15) & $-2.19\,(16)$ & 0.443\,(48) & 3.76\,(28) & 2.017\,(97)   \\
5 & 2.15\,(14) & $-2.16\,(16)$ & 0.432\,(47) & 3.64\,(27) & 1.948\,(91)   \\
6 & 2.20\,(14) & $-2.20\,(16)$ & 0.445\,(48) & 3.73\,(27) & 1.990\,(93)  \\[1ex]
7 & 2.19\,(10) & $-2.19\,(12)$ & 0.443\,(37) & 3.70\,(16) & 1.976\,(93)  \\[1ex]
phys. & 2.168\,(93) & $-2.18\,(10)$ & 0.436\,(29) & 3.65\,(15) & 1.945\,(76)   \\
\hline\hline\\[-1ex]& \multicolumn{5}{c}{$\langle B_d | O_n | \overline{B}_d\rangle_\msb^{(m_b)}/(f_{B_d}M_{B_d})^2$} \\[1ex]
set & $O_1$ & $O_2$ & $O_3$ & $O_4$ & $O_5$ \\
\hline\hline
1 & 2.07\,(21) & $-2.12\,(22)$ & 0.438\,(59) & 3.57\,(42) & 1.89\,(14)   \\
2 & 2.10\,(22) & $-2.13\,(22)$ & 0.421\,(60) & 3.77\,(44) & 2.04\,(15)   \\
3 & 2.06\,(21) & $-2.10\,(21)$ & 0.398\,(58) & 3.77\,(43) & 1.98\,(14)  \\[1ex]
4 & 2.20\,(16) & $-2.15\,(16)$ & 0.403\,(48) & 3.93\,(30) & 2.11\,(11)   \\
5 & 2.16\,(15) & $-2.06\,(15)$ & 0.396\,(48) & 3.70\,(28) & 1.965\,(98)   \\
6 & 2.11\,(16) & $-2.17\,(17)$ & 0.447\,(52) & 3.87\,(30) & 2.04\,(11)  \\[1ex]
7 & 2.20\,(12) & $-2.14\,(13)$ & 0.413\,(39) & 3.83\,(19) & 2.03\,(11)  \\[1ex]
phys. & 2.15\,(11) & $-2.06\,(11)$ & 0.400\,(30) & 3.82\,(18) & 2.015\,(92)   \\
\end{tabular}
\end{ruledtabular}

\end{table}

\begin{figure}
\includegraphics{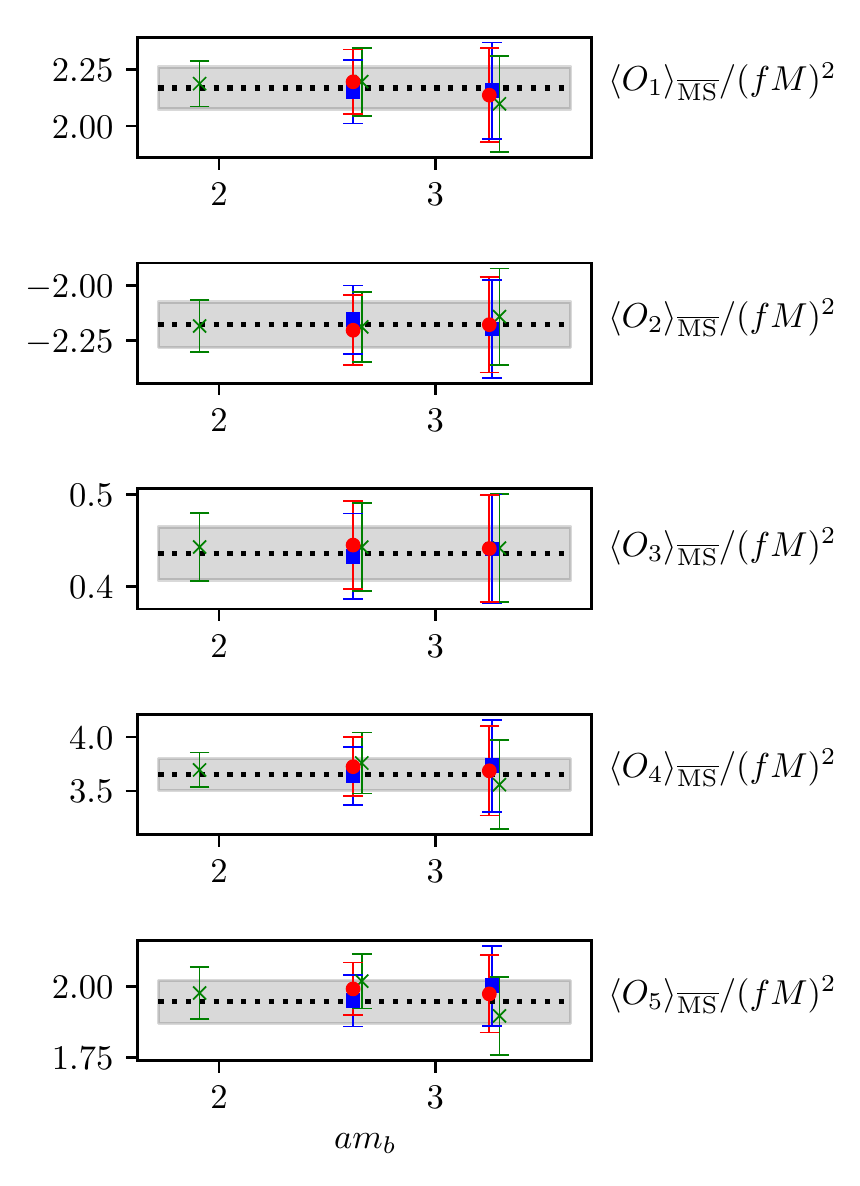}
    \caption{\label{fig:OnMSB} Comparison of the
    $\langle B_s | O_n | \overline{B}_s\rangle_\msb^{(m_b)}/(f_{B_s}M_{B_s})^2$
    values
    from individual  configurations sets (colored data points)
    with the final extrapolated values (gray bands and dotted lines) for each
    4-quark operator. Errors shown include correlated 
uncertainties from  operator normalisation and lattice spacing effects as discussed 
in the text. 
 The data are plotted versus $am_b$, falling into
    three groups corresponding to lattice spacings of 0.09, 0.12, and 0.15\,fm.
    Results are shown for three different values of light-quark mass
    $m_l\equiv(m_u+m_d)/2$ corresponding to $m_l/m_s=1/5$ (green, $\times$s),
    $m_l/m_s=1/10$ (blue, boxes), and the physical mass (red, circles).
    The dotted lines show the extrapolated values, while the gray bands show
    the $\pm1\,\sigma$ uncertainty in those values. The analogous figure
    for $B_d$~mesons is very similar.
    }
\end{figure}

\begin{figure}
\includegraphics{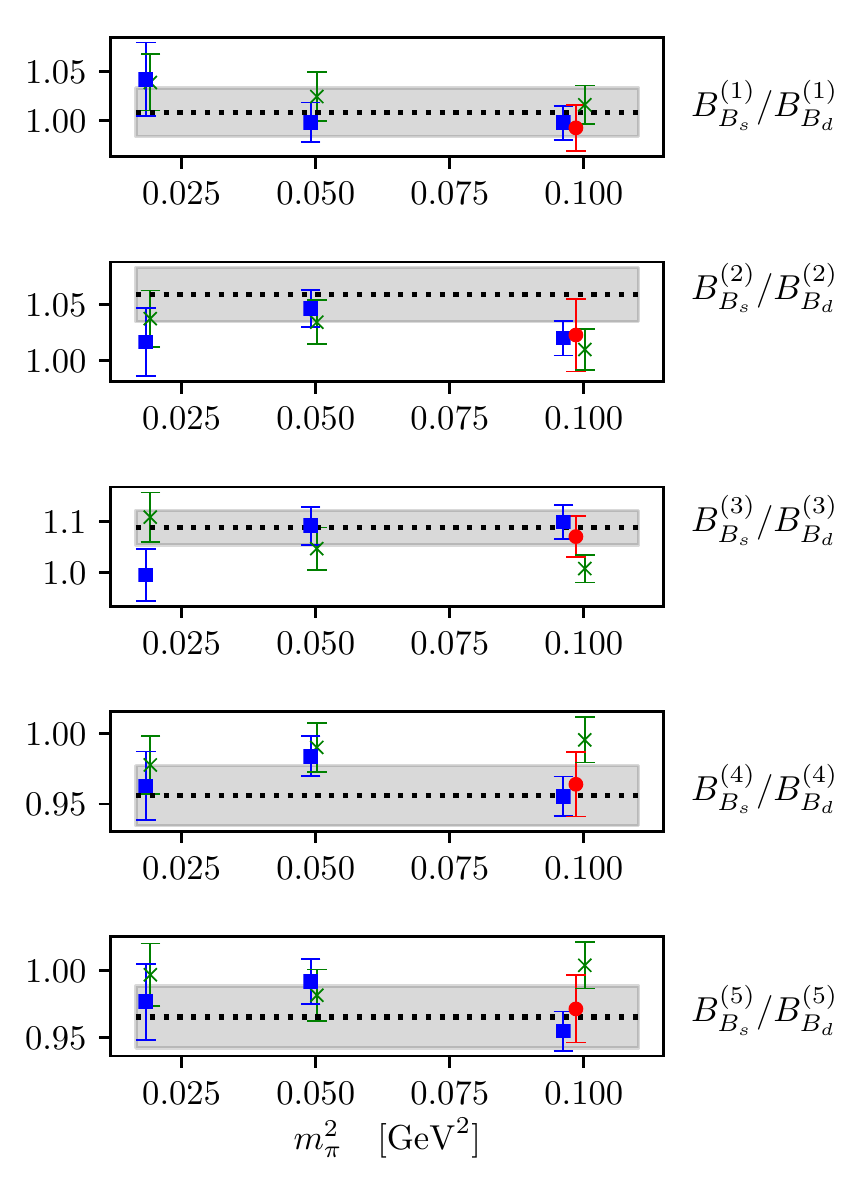}
    \caption{\label{fig:BsBd} Comparison of the ratio of bag parameters
    $B_{B_s}^{(n)}/B_{B_d}^{(n)}$
    from individual  configurations sets (colored data points)
    with the final extrapolated values (gray bands and dotted lines) for each
    4-quark operator. The data are plotted versus $m_\pi^2$, falling into
    three groups corresponding approximately to the physical value, 2.7~times
    the physical value, and 5.4~times the physical value.
    Results are shown for three different lattice spacings
    corresponding approximately to 0.15\,fm (green, $\times$s),
    0.12\,fm (blue, boxes), and 0.09\,fm (red, circles).
    The dotted lines show the extrapolated values, while the gray bands show
    the $\pm1\,\sigma$ uncertainty in those values.
    }
\end{figure}

We convert the lattice expectation values into $\msb$~matrix elements
using \Eq{matching} (divided by $(f_{B_q}M_{B_q})^2$).
Our results are listed in Table~\ref{tab:OnMSB}.
In addition to the statistical errors from the simulation and the
(negligible) errors in the $z_{ij}$, we include uncertainties
(for each entry in the table) coming from three additional sources:
\begin{itemize}
    \item $\mathcal{O}(\alpha_s^2)$: We estimate this uncertainty to be twice
    (to be conservative)
    $\alpha_s$ times the magnitude of the $\mathcal{O}(\alpha_s)$ correction
    we include for each of our three lattice spacings. (These corrections
    are correlated between configuration sets with similar lattice spacings.)

    \item $\mathcal{O}({\alpha_s\Lambda_\mathrm{QCD}/m_b})$:
    $\mathcal{O}(\Lambda_\mathrm{QCD}/m_b)$ corrections have been
    measured for the temporal axial-vector current
    and found to be~5\% of the leading-order contribution~\cite{Dowdall:2013tga}.
    This suggests that $\mathcal{O}(\Lambda_\mathrm{QCD}/m_b)$ corrections,
    which are included in our simulations,
    are~10\% for the 4-quark operators. We account for the $\mathcal{O}(\alpha_s)$
    radiative corrections
    to these terms by adding the following uncertainty to our results:
    \begin{equation}
    \label{eq:add-delta}
        \alpha_s\Big(
        c_{\alpha_s}^{n,0} + c_{\alpha_s}^{n,1} \delta_{a} + c_{\alpha_s}^{n,2} \delta^2_{a}
        \Big)
        \frac{\langle O_n\rangle_\msb}{(fM)^2}
    \end{equation}
    where each $c_{\alpha_s}^{n,i} = 0\pm0.1$ and
    \begin{equation}
        \delta_{a} \equiv  (am_b - 2.6)/1.4
    \end{equation}
    allows for variation in the coefficients between the lattice spacings.
    ($\delta_a$ is defined to vary from~$-1/2$ to~$1/2$ over our mass
    range; see \cite{Dowdall:2011wh} for more details.)

    \item $\mathcal{O}\big(\alpha_s (a\Lambda_\mathrm{QCD})^2,(a\Lambda_\mathrm{QCD})^4\big)$: The
    NRQCD and HISQ actions we use in the simulation are highly corrected.
    In particular there are no tree-level $a^2$~errors in either. We
    account for $a^2\alpha_s$ errors by adding the following uncertainty
    to our results:
    \begin{align}
        &\alpha_s \big(a \Lambda_\mathrm{QCD}\big)^2\Big(
        c_{a^2}^{n,0} + c_{a^2}^{n,1} \delta_{a} + c_{a^2}^{n,2} \delta^2_{a}
        \Big)
        \frac{\langle O_n\rangle_\msb}{(fM)^2}
        \nonumber \\
        &+ \big(a \Lambda_\mathrm{QCD}\big)^4\Big(
        c_{a^4}^{n,0} + c_{a^4}^{n,1} \delta_{a} + c_{a^4}^{n,2} \delta^2_{a}
        \Big)
        \frac{\langle O_n\rangle_\msb}{(fM)^2}
        \label{eq:a2-errors}
    \end{align}
    where each $c_{a^2}^{n,i}=0\pm1$, each
      $c_{a^4}^{n,i}=0\pm1$,
    $\Lambda_\mathrm{QCD}=0.5$\,GeV is the QCD scale,
    and again the $\delta_{a}$
    terms allow for variation between different lattice spacings.
\end{itemize}

The final entries (``phys.'') in both parts of
Table~\ref{tab:OnMSB} are our final results at the physical values of
the light-quark masses after our chiral fit.
We use chiral perturbation theory to
combine the values obtained on the different configuration
sets with different light quark masses;
see Appendix~\ref{sec:chifits} for details. Figure~\ref{fig:OnMSB} compares
our final values for $B_s$~mesons with the results from individual
configuration sets. These plots show that the dependence on lattice spacing and
light-quark mass is negligible compared with our uncertainties.
The analogous plot for $B_d$~mesons is very similar.

Adding uncertainties to the lattice results to allow for operator 
normalisation and lattice spacing effects, as we have done above, 
is equivalent to including them in our fit function with the coefficients 
treated as fit parameters; see the Appendix of~\cite{McNeile:2010ji}. 
The uncertainties that are included are correlated between lattice results on 
different sets through these coefficients.
Figure~\ref{fig:OnMSB} shows that, for example, the lattice spacing effects 
that we allow for through Eq.~(\ref{eq:a2-errors}) are overestimates 
of what is seen in the results, since the variation with lattice spacing of 
the central values is much smaller than the uncertainties shown on 
the coarser lattices. A further test of this is that 
omitting the results from the smallest 
lattice spacing (set 7) shifts our final central values by less than half 
a standard deviation and often much less.

\begin{table}
\caption{\label{tab:bagparam} $\msb$ bag parameters (\Eq{bagparam} with $\mu=m_b$) for the
five 4-quark operators. Results are given for both $B_s$~and
$B_d$~mesons, and for the ratios of bag parameters.
}

\begin{ruledtabular}
\begin{tabular}{cccccc}
 & $B_{B_q}^{(1)}(m_b)$ & $B_{B_q}^{(2)}(m_b)$ & $B_{B_q}^{(3)}(m_b)$ & $B_{B_q}^{(4)}(m_b)$ & $B_{B_q}^{(5)}(m_b)$ \\
\hline\hline
 $B_s$ & 0.813\,(35) & 0.817\,(43) & 0.816\,(57) & 1.033\,(47) & 0.941\,(38)  \\
 $B_d$ & 0.806\,(40) & 0.769\,(44) & 0.747\,(59) & 1.077\,(55) & 0.973\,(46)  \\
 $B_s/B_d$ & 1.008\,(25) & 1.063\,(24) & 1.092\,(34) & 0.959\,(21) & 0.967\,(23)  \\
\end{tabular}
\end{ruledtabular}

\end{table}

\begin{table}
\caption{\label{tab:errbudget} Percent errors coming from different sources
for the $B_s$~meson's bag parameters~$B_{B_s}^{(n)}$ and
$B_{B_s}^{(1)}/B_{B_d}^{(1)}$ (Table~\ref{tab:bagparam}). The total error
for each quantity is also shown. The error budgets for the $B_d$~meson's
bag parameters are
very similar. Systematic errors from
finite-volume, QED and strong-isospin breaking effects are estimated
to be below 0.1\% and hence negligible in Appendix~\ref{subsec:fvolsibqed}.
}

\begin{ruledtabular}
\begin{tabular}{lcccccc}
 & $B_{B_s}^{(1)}$ & $B_{B_s}^{(2)}$ & $B_{B_s}^{(3)}$ & $B_{B_s}^{(4)}$ & $B_{B_s}^{(5)}$ & $B_{B_s}^{(1)}/B_{B_d}^{(1)}$\\
\hline\hline
lattice data & 1.4 & 1.4 & 1.5 & 1.6 & 1.5 & 1.5 \\
$\eta_i^q$ & 0.0 & 2.3 & 2.3 & 2.1 & 1.2 & 0.0 \\
$\alpha_s^2$ terms & 2.1 & 2.9 & 5.2 & 1.9 & 1.5 & 0.1 \\
$\alpha_s \Lambda_\mathrm{QCD}/m_b$ terms & 2.9 & 2.8 & 2.9 & 2.8 & 2.7 & 0.0 \\
$(a\Lambda_\mathrm{QCD})^{2n}$ terms & 1.8 & 1.9 & 2.3 & 1.5 & 1.8 & 0.1 \\
$m_l$ extrapolation & 0.4 & 0.4 & 0.7 & 0.5 & 0.4 & 1.9 \\
\hline
Total & 4.3 & 5.3 & 7.0 & 4.6 & 4.1 & 2.5 \\
\end{tabular}
\end{ruledtabular}

\end{table}

Finally we convert our final results into bag parameters using
\Eq{bagparam}. The bag parameters are listed in Table~\ref{tab:bagparam}.
Despite the wide variation in values for $\langle O_n\rangle/(fM)^2$,
the bag parameters are within 30\% of~1.
This shows that the vacuum saturation approximation can be of some utility.

Figure~\ref{fig:BsBd} compares our final results for ratios of bag parameters
$B_{B_s}^{(n)}/B_{B_d}^{(n)}$ with results from the different configuration
sets. Results are plotted versus the value of $m_\pi^2$ used in each
simulations. Again there is very little variation with quark mass, with all
ratios within~5\% of~1. 
Our final results are shifted by less than half 
a standard deviation if we omit the data with the largest pion 
masses, and have errors that are 10-15\% larger.

The error budgets for the $B_s$ bag parameters are shown in
Table~\ref{tab:errbudget}. The dominant source of error comes from
uncalculated terms in perturbation theory ($\alpha_s^2$
and $\alpha_s \Lambda_\mathrm{QCD}/m_b$ terms). The sensitivity to these
terms depends on the operator. For example, it is particularly
high for $O_3$, because matrix elements for $O_3$ are a lot smaller than
those of $O_1$ (see Eq.~\ref{eq:etadef}) which are mixed in by Eq.~(\ref{eq:matching}).
 The error budgets
for $B_d$~mesons are almost identical to those for $B_s$, 
but have twice as large a contribution 
from statistical uncertainties in the lattice data. Almost
all of the uncertainties, and some of the statistical errors, cancel
in ratios of~$B_s$ to~$B_d$ meson bag parameters.

Matrix elements of the 4-quark mixing operators
can be obtained from the ratios in Table~\ref{tab:OnMSB}
given values for the decay constants and masses.
Note that the corresponding bag parameters for $O_{2\ldots 5}$ have
larger fractional errors than the ratios, and so should not be used for
this purpose. The larger errors result from uncertainties due to the
factors $\eta^q_i$ in the bag-parameter definition (see Table~\ref{tab:errbudget}
and Eq.~(\ref{eq:bagparam})).

\section{Discussion}
\label{sec:discussion}

\begin{figure}
  \includegraphics[width=\hsize]{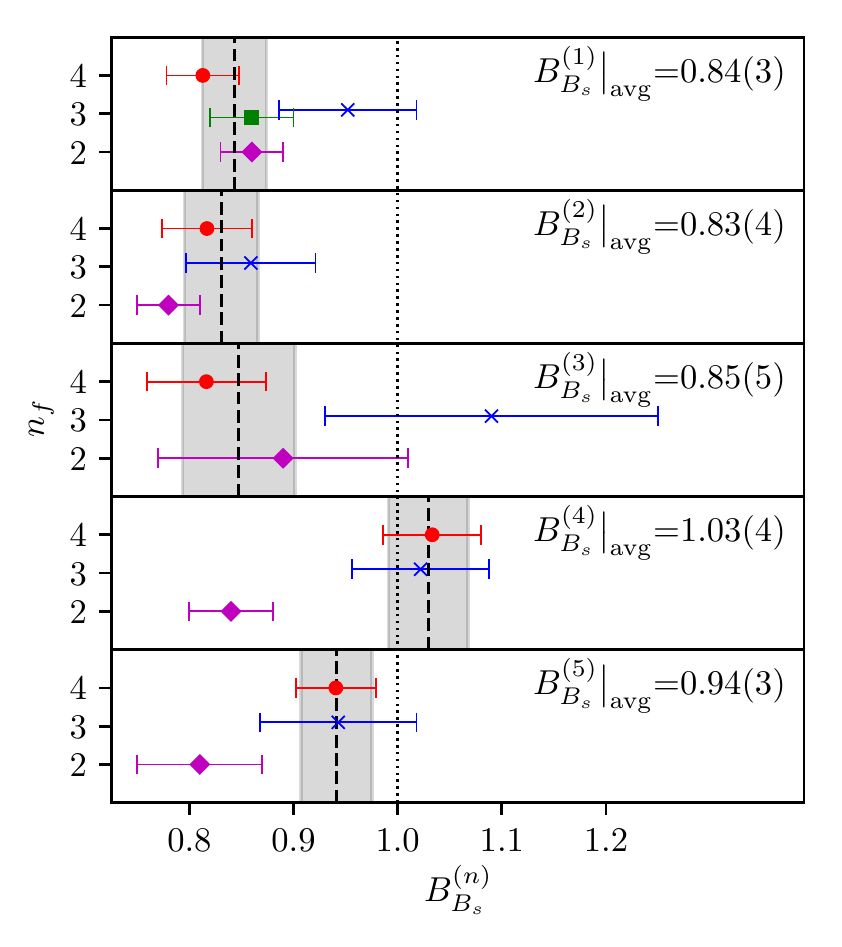}
\caption{\label{fig:Bscomp5} A comparison of our results
(red filled circles at $n_f=4$) to previous lattice QCD values
for the $B_s$ bag parameters $B_{B_s}(m_b)$ in the $\overline{\text{MS}}$
scheme
for all five SM and BSM operators.
Previous results come from the Fermilab/MILC collaboration
on $n_f=3$ gluon
field configurations (blue crosses)~\cite{Bazavov:2016nty} and
the ETM collaboration on $n_f=2$
gluon field configurations (purple filled diamonds)~\cite{Carrasco:2013zta}.
Note that the ETM results for $O_4$ and $O_5$ have been
converted to the definition of
the bag parameter given in Eq.~(\ref{eq:etadef}).
The filled green square at $n_f=3$ for the $O_1$ operator comes
from an earlier HPQCD calculation using NRQCD $b$ quarks~\cite{Gamiz:2009ku}.
The $n_f=2$ results are missing $s$ sea quarks, whose impact
cannot be estimated perturbatively (and no uncertainty is included
for this in the error bars). It is therefore unclear what
level of agreement to expect between these results and those
for $n_f=3$ and 4. Since we do not expect missing $c$ in the sea to
have a significant impact on the bag parameters~\cite{Bazavov:2016nty}
we can meaningfully compare $n_f=3$ and $n_f=4$. The grey bands are
the weighted average of our new results with
those of~\cite{Bazavov:2016nty}, and the average value
of the bag parameter $B^{(n)}_{B_s}(m_b)$ for each operator $O_n$
is indicated in that panel.
We include a vertical line at value 1.0 for comparison to the
vacuum saturation approximation.
}
\end{figure}

\begin{figure}
  \includegraphics[width=\hsize]{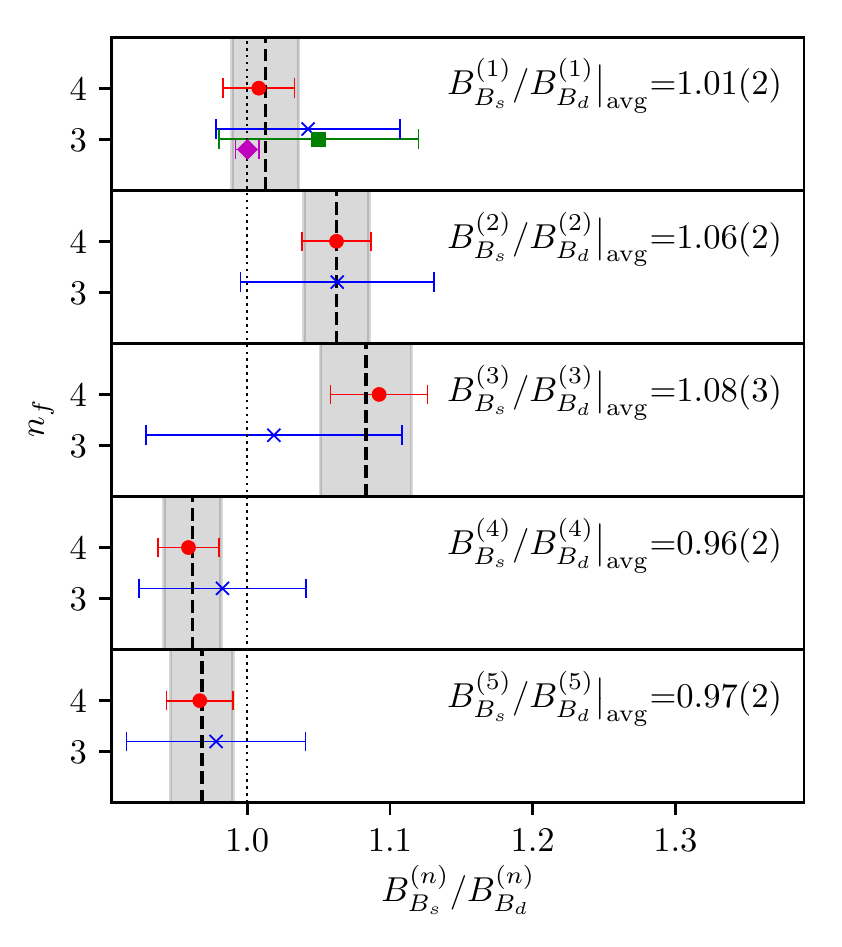}
\caption{\label{fig:BsBdcomp5} A comparison of our results
(red filled circles at $n_f=4$) to previous lattice QCD values
for the ratio of $B_s$ to $B_d$ bag parameters
for all five SM and BSM operators.
Previous results come from the Fermilab/MILC collaboration
on $n_f=3$ gluon
field configurations (blue crosses)~\cite{Bazavov:2016nty}
using their quoted correlations to reconstruct the ratio.
Since we do not expect missing $c$ in the sea to
have a significant impact on the bag parameters~\cite{Bazavov:2016nty}
we can meaningfully compare $n_f=3$ and $n_f=4$. The grey bands are
the weighted average of these two sets of results and the average value
for each operator is indicated in that panel.
For $O_1$ at $n_f=3$ we also show previous results
from HPQCD (green filled square)
using NRQCD $b$ quarks~\cite{Gamiz:2009ku} and RBC/UKQCD
(purple filled diamond) using domain-wall quarks with masses
of $m_c$ and above and extrapolating results
to the $b$ quark mass~\cite{Boyle:2018knm}.
We include a vertical line at value 1.0 to make clear which
ratios are above, and which below, this value.
}
\end{figure}

\subsection{Comparison to previous results}
\label{subsec:previous}

Our results for the bag parameters for all five SM and
BSM operators given in Table~\ref{tab:bagparam} are more
accurate than previous lattice
QCD results. This is for a number of reasons:
\begin{itemize}
\item
We work directly with the bag parameters
rather than the 4-quark operator matrix elements. The
bag parameters are expected from chiral perturbation theory
to have little
dependence on valence and
sea quark masses (see Appendix~\ref{sec:chifits}). This
expectation is borne out in our results and means that
we are able easily to combine results at both unphysical and
physical light quark masses.
\item We have
results for the physical light quark mass
at two values of the lattice spacing
improving control of the chiral extrapolation.
\item The gluon field configurations that we use include the
effect of $u$, $d$, $s$ and $c$ quarks in the sea and so we
do not have an uncertainty associated with missing flavours
of sea quarks (the Fermilab/MILC collaboration include a 2\%
uncertainty in their 4-quark operator matrix elements from missing
$c$ in the sea~\cite{Bazavov:2016nty}).
\end{itemize}

Figure~\ref{fig:Bscomp5} shows a comparison of our bag parameters
for the $B_s$ meson to those from~\cite{Bazavov:2016nty}
and~\cite{Carrasco:2013zta} (and also, for $O_1$, to~\cite{Gamiz:2009ku}).
The results from~\cite{Carrasco:2013zta}
include only $u$ and $d$ quarks in the sea and the uncertainty does
not include an estimate of the impact of missing $s$ sea quarks.
It is therefore not clear whether we should expect agreement between
these $n_f=2$ results and our $n_f=4$ results.
The fact that the $n_f=2$ purple diamonds from the ETM collaboration
are around 20\% below our results for $O_4$ and $O_5$ is reminiscent
of what is seen in kaon mixing. ETM use the RI-MOM renormalisation scheme
for the purple diamonds and it has been shown in
kaon mixing~\cite{Garron:2016mva} that the
use of the RI-MOM scheme (rather than RI-SMOM) for the equivalent 4-quark
operators has large systematic errors that push down the
value of the bag parameter. This may then be the main reason
(rather than a difference of $n_f$)
for the discrepancy with our results for $O_4$ and $O_5$,
but more work would be needed to be sure of this.  

The $n_f=3$ and $n_f=4$ results should be comparable because the
impact of missing $c$ quarks in the sea on the bag parameters is
expected to be very small~\cite{Bazavov:2016nty}.
Our new results agree within
$2\sigma$ in each case with~\cite{Bazavov:2016nty} but in every
case are more accurate. The largest discrepancy is for $B^{(1)}_{B_s}$
at $1.9\sigma$.

The weighted average
of our $n_f=4$ results and the $n_f=3$ results
from~\cite{Bazavov:2016nty} is given by
the grey band in the Figure and the value of that average is
given in each panel. We assume no correlations,
here and subsequently, between our
results and those of~\cite{Bazavov:2016nty} because they use
different actions for both the $b$ quark and the light quarks
and different gluon field configurations (with a different sea
quark action and generated with a different
Monte Carlo updating algorithm).

Figure~\ref{fig:BsBdcomp5} shows a comparison of the ratio of
bag parameters for $B_s$ to $B_d$ for each operator for our new
results and those of~\cite{Bazavov:2016nty}.
Our new results are a lot
more accurate, with 2--3\% total uncertainty. All of the ratios are
very close to 1, but there is a sign of a systematic trend for the
ratio for $O_2$ and $O_3$ to be above 1 and for $O_4$ and $O_5$ below 1.
This is not visible in the results of~\cite{Bazavov:2016nty} but does start
to emerge with the improved accuracy of our results.
This is in general agreement with the results from using
sum rules in~\cite{Grozin:2017uto,King:2019lal}.
We also include in Figure~\ref{fig:BsBdcomp5} results for $O_1$ from
HPQCD~\cite{Gamiz:2009ku} and RBC/UKQCD~\cite{Boyle:2018knm}.
The RBC/UKQCD result has a 1\% uncertainty.

\subsection{Derived quantities}
\label{subsec:derived}

Our results for the bag parameters can be combined with results for
the $B$ and $B_s$ decay constants to give values for
the 4-quark operator matrix elements using Eq.~(\ref{eq:red-mat-element})
and our results in Table~\ref{tab:OnMSB}.
For this we use the most accurate current lattice QCD results
obtained on gluon field configurations including $u/d$, $s$ and $c$
quarks in the sea. These have been obtained by the Fermilab/MILC
collaboration using the HISQ action for all quarks~\cite{Bazavov:2017lyh}.
This `heavy-HISQ' approach, pioneered by
HPQCD~\cite{McNeile:2011ng, McNeile:2012qf}, uses pseudoscalar meson
2-point correlators
that combine heavy and light quark propagators calculated with
multiple heavy quark masses, $am_h$, at multiple values of the lattice spacing.
$m_h$ reaches the $b$ quark mass for
$am_h < 1$ for lattice spacing values $a < 0.045$ fm.
Since the HISQ action has very small discretisation errors
by design,  a fit to the $m_h$- and $a-$dependence is possible
that allows the continuum $m_h$-dependence of the decay constant
to be reconstructed. It can then be evaluated at the $b$ quark
mass to enable the $B$ and $B_s$ decay constants to be determined.
Note that the correlators can be absolutely normalised in this case
and so there is no normalisation uncertainty.

Fermilab/MILC obtain the values $f_{B_d}=0.1905(13)$ GeV,
$f_{B_s}=0.2307(13)$ GeV and $f_{B_s}/f_{B_d}=1.2109(41)$\footnote{Our
results obtained on $n_f=2+1+1$ gluon field configurations from
NRQCD-HISQ calculations~\cite{Dowdall:2013tga, Hughes:2017spc} agree with these numbers but are less accurate.}.
Note that we use the decay constant for the neutral $B_d$ meson
(not the $B_u$), which is the appropriate choice here.
Our bag parameters
are calculated for a light quark~$l$ corresponding to the average
of $u$ and $d$. Our results show (comparing those for $B_s$ with
those for $B$) that any difference between bag parameters
for $B_l$ and $B_d$ will be much smaller than our uncertainties.
This is not true for the decay constants, where the differences
are significant~\cite{Bazavov:2017lyh}.

For the SM phenomenology to be determined from our results for
the matrix elements of $\mathcal{O}_1$ it is convenient to
convert our results from the $\overline{\text{MS}}_{\text{NDR}}$
scheme to the renormalisation-group-invariant quantities
$\hat{B}^{(1)}_{B_q}$. The conversion is given by
\begin{equation}
\label{eq:rgiconv}
\hat{B}^{(1)}_{B_q} = c_{\text{RGI}}B^{(1)}_{B_q}(m_b)\, .
\end{equation}
The matching factor $c_{\text{RGI}}$ is calculated to
two-loops in perturbative
QCD and we take
$c_{\text{RGI}}=1.5158(36)$~\cite{Bazavov:2016nty}.
This corresponds to the result for $n_f=5$ active flavours in
the sea and $\overline{\alpha}_s(M_Z)=0.1185(6)$.
Our bag parameters are obtained at scale
$m_b$ for 4 flavours of quarks in the sea. The impact of
missing $b$ quarks in the sea, however, should be negligible
both for the bag parameters and the resulting 4-quark operator
matrix elements.  A power-counting estimate of such effects
would give a relative contribution of
$\alpha_s(\Lambda_{\text{QCD}}/2m_b)^2$, which is below 0.1\%.

Our results for the RGI bag parameters for $\mathcal{O}_1$ are
then:
\begin{eqnarray}
\label{eq:bhatres}
\hat{B}^{(1)}_{B_s} &=&  \bhats \\
\hat{B}^{(1)}_{B_d} &=&  \bhatd \nonumber \\
\frac{\hat{B}^{(1)}_{B_s}}{\hat{B}^{(1)}_{B_d}} &=&  \BBsBBdratio \, .\nonumber
\end{eqnarray}
The ratio of RGI bag parameters is of course the same as that of
the $\overline{\text{MS}}$ bag parameters.
Combined with the decay constant results from~\cite{Bazavov:2017lyh}
we obtain
\begin{eqnarray}
\label{eq:fsqbhatres}
f_{B_s}\sqrt{\hat{B}^{(1)}_{B_s}} &=&  \fsqrtbhats \\
f_{B_d}\sqrt{\hat{B}^{(1)}_{B_d}} &=&  \fsqrtbhatd \nonumber \\
\xi &=& \xifinal \nonumber
\end{eqnarray}
where $\xi$ is the ratio of the two results above it. We form
$\xi$ by combining the result for $f_{B_s}/f_{B_d}$
from~\cite{Bazavov:2017lyh} with our results
for $B^{(1)}_{B_s}/B^{(1)}_{B_d}$, taking advantage
of the correlations that reduce uncertainties in each of these ratios.
Note that in combining the decay constant
and bag parameter results we add relative uncertainties in quadrature.
We expect no significant correlation between the two sets of results
because they use a different heavy quark action
and, even though both results
use $n_f=2+1+1$ gluon field configurations, there is little overlap
in the ensembles used. The error budgets in the two cases show that
the key sources of uncertainty are not the same.
The uncertainties in the combinations above
are dominated by the uncertainties in our bag parameters and their
ratio in Eq.~(\ref{eq:bhatres}) because the decay constant
results are now so accurate.

\begin{figure}
  \includegraphics[width=\hsize]{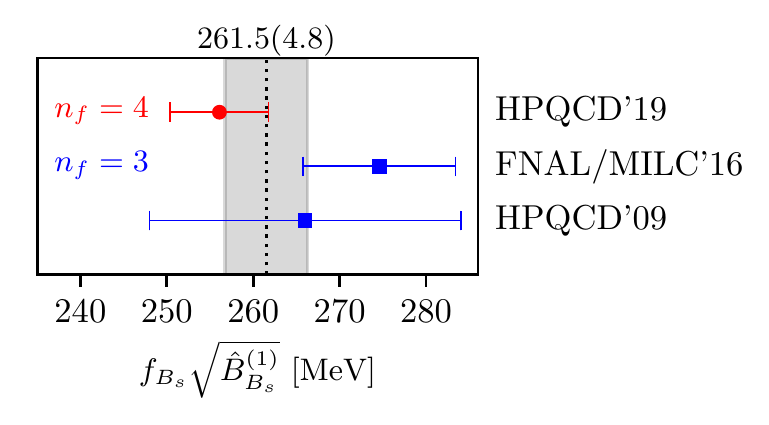}
\caption{\label{fig:fsqrtBcomp} A comparison of our results
(red filled circles at $n_f=4$) to previous lattice QCD values
for the combination of decay constant and square root of bag
parameter $f_{B_s}\sqrt{\hat{B}^{(1)}_{B_s}}$.
Previous results (blue filled squares) come from the
Fermilab/MILC
collaboration~\cite{Bazavov:2016nty} and
from HPQCD~\cite{Gamiz:2009ku} on $n_f=3$ gluon field configurations.
The Fermilab/MILC results include a 1\% uncertainty for missing
$c$ in the sea.
The grey band is
the weighted average of our new results and those of~\cite{Bazavov:2016nty}
and the new lattice QCD average value is quoted at the top.
}
\end{figure}

\begin{figure}
  \includegraphics[width=\hsize]{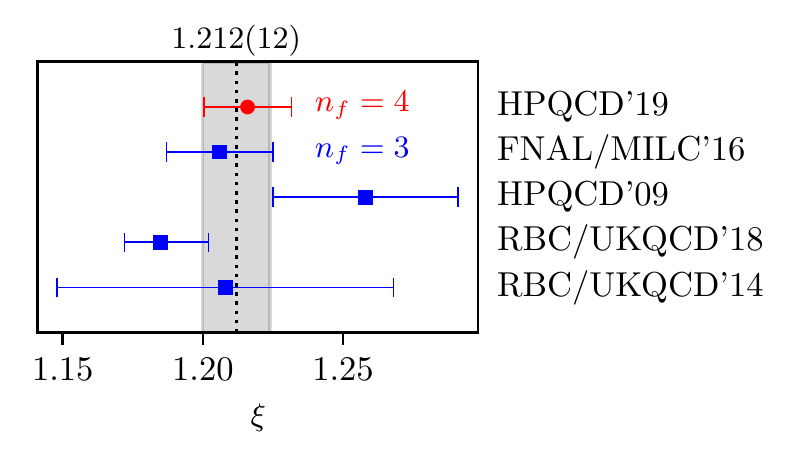}
\caption{\label{fig:xicomp} A comparison of our results
(red filled circles at $n_f=4$) for $\xi$, defined in
Eq.~(\ref{eq:fsqbhatres}), to previous lattice QCD values
for $n_f=3$ (filled blue squares).
Previous results come from the
Fermilab/MILC
collaboration~\cite{Bazavov:2016nty} and
from HPQCD~\cite{Gamiz:2009ku} using calculations at the physical
$b$ quark mass. Results are also shown from RBC/UKQCD
using domain-wall quarks and extrapolating to the $b$ from
the $c$ quark region and above~\cite{Boyle:2018knm} and
using static (infinitely massive) $b$ quarks~\cite{Aoki:2014nga}.
The grey band is
the weighted average of our new results and those of~\cite{Bazavov:2016nty}
with the result for the average quoted above it.
}
\end{figure}

Figure~\ref{fig:fsqrtBcomp} compares our new result for
$f_{B_s}\sqrt{\hat{B}_{B_s}}$ from Eq.~(\ref{eq:fsqbhatres})
to previous lattice QCD results on $n_f=3$ gluon field configurations
from Fermilab/MILC~\cite{Bazavov:2016nty} and HPQCD~\cite{Gamiz:2009ku}.
The Fermilab/MILC results include an uncertainty for missing
$c$ in the sea in their calculation.
The difference between the central value of our new result and
that of Fermilab/MILC is $1.8\sigma$.
Because the systematic uncertainties are correlated between our
results and those of~\cite{Gamiz:2009ku} we do not include the
previous HPQCD results in the new lattice QCD $n_f=3/n_f=4$ average,
shown by the grey band in Figure~\ref{fig:fsqrtBcomp}. The average value is
shown above the grey band.

Figure~\ref{fig:xicomp} compares lattice QCD results
for the ratio $\xi$ defined
in Eq.~(\ref{eq:fsqbhatres}) on $n_f=3$ gluon field configurations
with our new result here using $n_f=4$. There is good agreement between
the lattice QCD results with the most recent (including our new result here)
having total uncertainties at the level of 1.5\%.
The result of averaging our new result with that of~\cite{Bazavov:2016nty}
(both results being obtained at the physical $b$ quark mass) is
given by the grey band with the average value quoted above it.

\subsection{$\Delta M$}
\label{subsec:deltaM}

The phenomenon of neutral $B$-meson oscillations is now
well-established experimentally (for recent results
see~\cite{Abe:2004mz, Aubert:2005kf, Abazov:2006qp, Aaij:2016fdk, Abulencia:2006ze, Aaij:2011qx, Aaij:2013mpa, Aaij:2013gja, Aaij:2014zsa}),
with an oscillation frequency that
is set by the mass difference between the two eigenstates.
The current experimental average values~\cite{Tanabashi:2018oca}
for the $B_s$ and
$B_d$ systems are:
\begin{eqnarray}
\label{eq:deltamexp}
\Delta M_{s, \text{expt}} = 17.757(21) {\text{ps}}^{-1} \\
\Delta M_{d,\text{expt}} = 0.5065(19) {\text{ps}}^{-1} \nonumber
\end{eqnarray}
combining statistical and systematic errors in quadrature.

In the SM $\Delta M$ is given by
\begin{equation}
\label{eq:deltamSM}
\Delta M_q = \frac{G_F^2M_W^2M_{B_q}}{6\pi^2}S_0(x_t)\eta_{2B}\left|V_{tq}^*V_{tb}\right|^2f^2_{B_q}\hat{B}^{(1)}_{B_q} \, .
\end{equation}
Here $S_0$ is the Inami-Lim function~\cite{Inami:1980fz} which
describes electroweak corrections
and has argument $x_t=m_t^2/M_W^2$. The top quark mass to be used here is
in the $\overline{MS}$ scheme,
$\overline{m}_t(\overline{m}_t)$~\cite{Buchalla:1995vs}.
Taking the current average~\cite{Tanabashi:2018oca} of direct experimental
measurements~\cite{Khachatryan:2015hba, Aaboud:2016igd, Sirunyan:2018gqx}
of the top quark mass (172.9(4) GeV) as the pole mass,
gives $\overline{m}_t(\overline{m}_t)$ = 163.07(38) GeV
using the 4-loop expressions in~\cite{Marquard:2015qpa}.
Evaluating the Inami-Lim function then gives:
$S_0(4.116(19))=2.313(8)$.
The QCD correction factor, $\eta_{2B}$, is given at
next-to-leading order in~\cite{Buras:1990fn}.
We take $\eta_{2B}$=0.55210(62)~\cite{Bazavov:2017lyh},
again calculated with $n_f=5$.

The CKM elements $V_{tq}$ and $V_{tb}$ can be derived in the SM by
assuming that the CKM matrix is unitary and determining other CKM
elements in the same rows or columns from the comparison of
theory and experiment~\cite{Charles:2004jd, Bona:2006ah, CKMfitter, UTfit}. For Eq.~(\ref{eq:deltamSM})
it is important to use values for $V_{tq}$ that did not include
$\Delta M_q$ itself in their determination. So we use the results
from CKMfitter for the case where only tree-level processes were used
in the determination. This gives~\cite{CKMfitter}
\begin{eqnarray}
\label{eq:ckmvtq}
\left| V_{ts} \right|_{\text{CKMfitter, tree}} &=& \left(41.69^{+0.39}_{-1.45}\right) \times 10^{-3} \\
\left| V_{td} \right|_{\text{CKMfitter, tree}} &=& \left(9.08^{+0.23}_{-0.45}\right) \times 10^{-3} \nonumber \\
\left| V_{td} / V_{ts} \right|_{\text{CKMfitter, tree}} &=& 0.2186^{+0.0049}_{-0.0059}  \nonumber \\
\left| V_{tb} \right|_{\text{CKMfitter, tree}} &=& 0.999093^{+0.000064}_{-0.000018} \, .\nonumber
\end{eqnarray}
The ratio $\left|V_{td}/V_{ts}\right|_{\text{CKMfitter, tree}}$ is derived
from the CKMfitter results for $A$, $\lambda$, $\overline{\rho}$
and $\overline{\eta}$ using the formulae in~\cite{Charles:2004jd}. The
central value differs
slightly from the ratio of the two numbers above.

The final terms in Eq.~(\ref{eq:deltamSM})
parameterise the hadronic contribution
to $\Delta M$ through the matrix element of
the appropriate 4-quark operator, $\mathcal{O}_1$.
Our results for $f_{B_q}^2\hat{B}^{(1)}_{B_q}$ are
given in Eq.~(\ref{eq:fsqbhatres}).

Putting all these pieces together we obtain predictions for the mass
differences for neutral $B_s$ and $B_d$ eigenstates of
\begin{eqnarray}
\label{eq:ourdeltaM}
\Delta M_{s,\text{SM}} &=&  17.59(^{+0.33}_{-1.22})(0.78)\,\mathrm{ps}^{-1}      \\
\Delta M_{d, \text{SM}} &=&  0.555(^{+28}_{-55})(29)\,\mathrm{ps}^{-1}      \nonumber \\
\left(\frac{\Delta M_d}{\Delta M_s}\right)_{\text{SM}} &=&  0.0318(^{+14}_{-17})(8)  \, ,    \nonumber
\end{eqnarray}
where the first error in each case is from the CKM matrix elements and the second
error is primarily from the lattice analyses.
These results agree well with the experimental values from Eq.~(\ref{eq:deltamexp})
--- the largest discrepancy is $1.7\sigma$ for the ratio of $\Delta M$
values --- but they have much larger uncertainty.

\subsection{$V_{ts}$ and $V_{td}$}
\label{subsec:ckm}

Because the experimental values for $\Delta M_q$ are so
accurate, a better approach to understanding the
implications of our improved lattice
QCD results for
the relevant hadronic matrix elements
is to turn the analysis of the previous subsection on its head.
That is, to use our results and the experimental values for
$\Delta M_q$ to determine values
for $|V_{ts}|$ and $|V_{td}|$ from Eq.~(\ref{eq:deltamSM}) (taking a value for
$V_{tb}$ from Eq.~(\ref{eq:ckmvtq})~\cite{CKMfitter}).
$|V_{ts}|$ and $|V_{td}|$ obtained this way can then be compared to other determinations
that make use of CKM unitarity as a test of that unitarity.

The ratio of $|V_{ts}|$ to $|V_{td}|$ can be obtained more
accurately than the separate CKM elements because this makes
use of the hadronic parameter $\xi$ (Eq.~(\ref{eq:fsqbhatres}))
in which a lot of the lattice QCD
uncertainties cancel (see Section~\ref{subsec:previous}).

Our results are
\begin{eqnarray}
\label{eq:vresults}
\left| V_{td} \right| &=& \vtdfinal \\
\left| V_{ts} \right| &=& \vtsfinal \nonumber \\
\left| V_{td} \right|/\left|V_{ts}\right| &=& \vratio \, .\nonumber
\end{eqnarray}

\begin{figure}
  \includegraphics[width=\hsize]{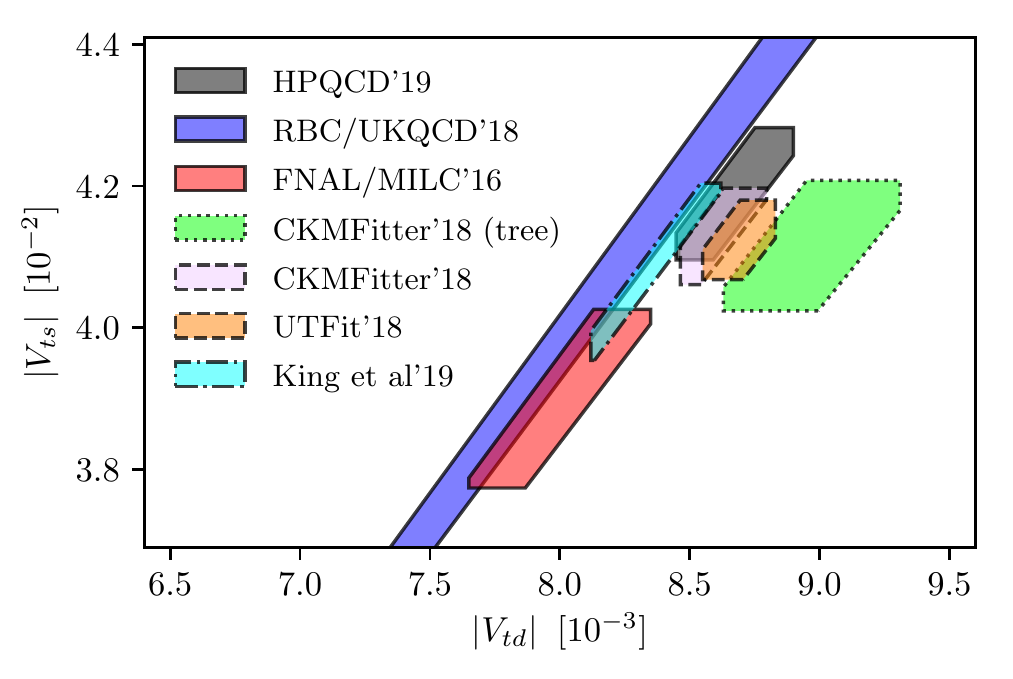}
\caption{\label{fig:VtdVts} A comparison of $\pm 1 \sigma$
constraints on $V_{ts}$ and $V_{td}$ from
experimental results on $B_s$ and $B_d$ oscillation frequencies
compared to SM calculations. This is an update of Figure 7
in~\cite{King:2019lal} to include the results presented here.
The lattice QCD constraints shown come from: this paper, dark grey;
\cite{Bazavov:2016nty}, red;~\cite{Boyle:2018knm}, light blue, $|V_{ts}|/|V_{td}| $ ratio only. The light blue lozenge is from sum rules~\cite{King:2019lal}.
The lozenges with dashed boundaries include a full unitarity triangle
fit: light pink is from CKMfitter~\cite{Charles:2004jd, CKMfitter}
and orange from UTFit~\cite{Bona:2006ah, UTfit}. The green lozenge
with dotted boundary is the result of a unitarity triangle fit
for tree-level processes only from CKMfitter.
}
\end{figure}

Figure~\ref{fig:VtdVts} plots the $\pm 1 \sigma$ constraints on
$|V_{td}|$, $|V_{ts}|$
and their ratio from our results as the dark grey lozenge.
Results determined by other lattice QCD
calculations~\cite{Bazavov:2016nty, Boyle:2018knm} are also shown
along with a recent determination using sum rules~\cite{King:2019lal}.
Also shown as light pink and orange lozenges
are results from fits to the CKM unitarity triangle
using results from many different processes~\cite{CKMfitter, UTfit}.
Particularly relevant
here is the green lozenge which results from a unitarity triangle fit
that includes tree-level processes only~\cite{CKMfitter}, and
therefore not $B_s/B_d$ oscillations.
Tension between results derived from $\Delta M_q$ (as here) and the
results derived from tree-level processes and unitarity would imply
the existence of new physics in loop processes.

The Fermilab/MILC results (red lozenge in Figure~\ref{fig:VtdVts})
highlighted an approximately $2.0\sigma$ tension between their values for
$V_{ts}$ and $V_{td}$ and those from unitarity fits.
See~\cite{Blanke:2016bhf, DiLuzio:2017fdq} for examples of
the possible implications of this.

Our results show no such tension. Our values for $V_{ts}$ and $V_{td}$
separately agree with the \{CKMfitter, tree\} results in Eq.~(\ref{eq:ckmvtq})
within $1\sigma$ and the difference in the ratio amounts to $1.8\sigma$.
This limits the scope for new physics in loop-induced processes.
However,
our ratio for $|V_{td}|/|V_{ts}|$ joins the systematic trend of the
previous results shown in Figure~\ref{fig:VtdVts} in being
below that of \{CKMFitter, tree\}.

\subsection{$B_q \rightarrow \mu^+\mu^-$ decay}
\label{subsec:Bmumu}

The rare decays $B_q \rightarrow \mu^+\mu^-$ have very small
branching fractions in the SM since they proceed through $W$ box
diagrams and $Z$ penguins and are helicity-suppressed.
New physics might then be seen if the experimental and SM branching
fractions can be determined to be different to sufficient accuracy.

The hadronic parameter that enters the SM branching fraction
is the $B_q$ meson decay constant~\cite{Bobeth:2013uxa} but
it appears along with the CKM elements $\left|V^*_{tq}V_{tb}\right|$.
The uncertainty in the value of the appropriate CKM element is now
the largest uncertainty in the value of the
SM branching fraction~\cite{Bazavov:2017lyh}.

An alternative method for determining the branching fraction is
to take a ratio to $\Delta M$~\cite{Buras:2003td}. In the SM
(and extensions with minimal flavour violation) the CKM elements
cancel out of this ratio. The decay constant also cancels and
the hadronic parameter that remains in the ratio is the bag parameter.

The formula for the time-averaged branching
fraction~\cite{DeBruyn:2012wk}, as measured
in the experiment, is then given in the SM by
\begin{eqnarray}
\label{eq:bmumu}
\frac{{\text{Br}}(B_q \rightarrow \ell^+\ell^-)}{\Delta M_q} &=& \\
&& \hspace{-4.0em}\frac{3 G_F^2M_W^2m_{\ell}^2}{\pi^3}\tau_{B_q^H}\sqrt{1-\frac{4m^2_{\ell}}{M_{B_q}^2}}\frac{|C_A(\mu_b)|^2}{S_0(x_t)\eta_{2B}\hat{B}^{(1)}_{B_q}} \, .\nonumber
\end{eqnarray}
Here $C_A(\mu_b)$ includes electroweak and QCD corrections and
is given for $\mu_b$= 5 GeV in~\cite{Bobeth:2013uxa}.
We use $C_A(\mu_b)$ = 0.4694(36)~\cite{Bazavov:2016nty}.
The lifetime $\tau_{B_q^H}$ that appears in this formula is
that of the heavy neutral eigenstate~\cite{DeBruyn:2012wk}. For the
$B_d$ this can be taken as the average lifetime, 1.520(4) ps~\cite{Amhis:2016xyh,hflav}
but for the $B_s$ there is a measured difference of lifetimes and the
heavy eigenstate has the longer lifetime, 1.615(9) ps~\cite{Amhis:2016xyh, hflav}.
Values for $S_0(x_t)$ and $\eta_{2B}$ are given in Section~\ref{subsec:ckm}.

\begin{figure}
  \includegraphics[width=\hsize]{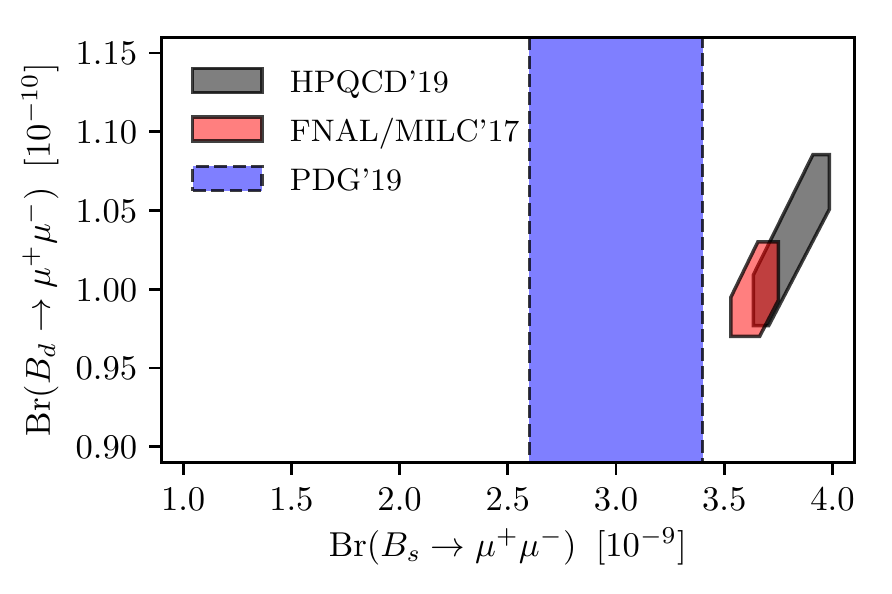}
\caption{\label{fig:BrBsBd}
A comparison of the SM branching fractions for $B_s$
and $B_d$ to decay to $\mu^+\mu^-$ from lattice QCD results with the current
experimental measurements. The grey lozenge shows results from
our calculation here of the bag parameters and a ratio to experimental
results for $\Delta M_q$ (Eqs.~(\ref{eq:ourbmumu}) and~(\ref{eq:usbmumurat})).
The red lozenge shows results from a Fermilab/MILC calculation of
$B_d$ and $B_s$ decay constants, combined with input CKM
elements~\cite{Bazavov:2017lyh}. The blue band shows the current experimental
average for ${\text{Br}}(B_s \rightarrow \mu^+\mu^-)$~\cite{Tanabashi:2018oca}; only an upper bound
exists for ${\text{Br}}(B_d \rightarrow \mu^+\mu^-)$.
}
\end{figure}

Using our results for the bag parameters for $\mathcal{O}_1$
given in Eq.~(\ref{eq:bhatres}) and the experimental values
for $\Delta M$ in Eq.~(\ref{eq:deltamexp}) we obtain the following
values for the branching fractions:
\begin{eqnarray}
\label{eq:ourbmumu}
{\text{Br}}(B_s \rightarrow \mu^+\mu^-) &=& \Bsmumu \\
{\text{Br}}(B_d \rightarrow \mu^+\mu^-) &=& \Bdmumu \, .\nonumber
\end{eqnarray}

We can also obtain the ratio of branching fractions for $B_s$ and
$B_d$~\cite{Buras:2003td}
\begin{equation}
\label{eq:bmumurat}
\frac{{\text{Br}}(B_s \rightarrow \mu^+\mu^-)}{{\text{Br}}(B_d \rightarrow \mu^+\mu^-)} = \frac{\tau_{B_s^H}}{\tau_{B_d^H}}\frac{\hat{B}^{(1)}_{B_d}}{\hat{B}^{(1)}_{B_s}}\frac{\Delta M_s}{\Delta M_d} .
\end{equation}
Here we have dropped the terms in $m^2_{\mu}/M^2_{B_q}$ since they are negligible.
There is a lot of cancellation in this ratio, including of systematic
errors in the ratio of bag parameters (see our results in Eq.~(\ref{eq:bhatres})).
We obtain the result
\begin{equation}
\label{eq:usbmumurat}
\frac{{\text{Br}}(B_d \rightarrow \mu^+\mu^-)}{{\text{Br}}(B_s \rightarrow \mu^+\mu^-)} = \BdmumuBsmumu .
\end{equation}

Figure~\ref{fig:BrBsBd} shows our predictions
in the SM for the branching fractions
from Eqs.~(\ref{eq:ourbmumu}) and~(\ref{eq:usbmumurat}) as the grey lozenge.
The red lozenge
shows lattice QCD predictions~\cite{Bazavov:2017lyh} for the
branching fractions using the direct approach
where the hadronic parameter needed is the decay constant and this is combined with
input for the CKM elements $V_{tq}$ and $V_{tb}$, along with other factors.
The errors in the results from~\cite{Bazavov:2017lyh} are
dominated by uncertainties in the CKM elements, which
are taken from a global unitarity triangle
fit that includes both tree and loop-induced processes\footnote{Note that 
we include the constraint on the ratio from the July 2019 update of~\cite{Bazavov:2017lyh}.}.

Figure~\ref{fig:BrBsBd} shows good agreement between the
two lattice QCD predictions.
This reflects the fact that, as described in Section~\ref{subsec:ckm} our
results for the bag parameters yield CKM elements $|V_{ts}|$ and $|V_{td}|$ in
agreement with CKM unitarity determinations. Our results imply consistency of
the CKM matrix (within uncertainties) and hence the two approaches of using the
decay constants plus CKM elements or using the bag parameters and $\Delta M_q$
will agree.

Note that our results in Eq.~(\ref{eq:ourbmumu}) include
uncertainties in the parameters of Eq.~(\ref{eq:bmumurat}).
They do not include uncertainties from electromagnetic corrections
to the decay process. These are estimated to lead to
a reduction of 0.3--1.1\% in the muonic branching fractions
in~\cite{Beneke:2017vpq}. This is not significant given the current
uncertainties in our SM predictions, but will need to be addressed
as reduce uncertainties in future.

The blue band in Figure~\ref{fig:BrBsBd} shows the current experimental situation.
The decay $B_d \rightarrow \mu^+\mu^-$ has only been seen with $3\sigma$
significance~\cite{CMS:2014xfa}. Recent LHCb~\cite{Aaij:2017vad} and
ATLAS~\cite{Aaboud:2018mst} results give upper bounds to
the branching fraction of  $3.4 \times 10^{-10}$ and $2.1 \times 10^{-10}$ respectively.
These bounds are outside the range of Figure~\ref{fig:BrBsBd}.
For the branching fraction for $B_s \rightarrow \mu^+\mu^-$, the Particle
Data Group quotes an average value of $3.0(4) \times 10^{-9}$ using results
from ATLAS~\cite{Aaboud:2018mst}, CMS~\cite{Chatrchyan:2013bka} and
LHCb~\cite{Aaij:2017vad}. The $\pm 1 \sigma$ variation gives the width of the band
in Figure~\ref{fig:BrBsBd}.

Although no significant tension between experiment and the SM predictions is
visible in this figure, it does give encouragement that we are reaching a point
where further reductions in uncertainties will start give serious SM constraints,
at least for $B_s \rightarrow \mu^+\mu^-$.

\subsection{Contributions to $\Delta \Gamma$}
\label{subsec:deltagamma}

Another physical observable from neutral $B$ meson systems
is that of the decay width difference of the
eigenstates, $\Delta \Gamma$. This has been
measured for the $B_s$ at 13\% of the average width, but
only an upper limit exists for the $B_d$~\cite{Tanabashi:2018oca}.
The prediction for the width
differences in the SM is given in~\cite{Beneke:1998sy, Lenz:2006hd}
in terms of the matrix elements of several 4-quark operators.
We give results here for the matrix elements of
those labelled $R_0$, $R_1$ and $\tilde{R}_1$ in~\cite{Lenz:2006hd}.

$R_0$~\cite{Lenz:2006hd} is a combination of $O_1$, $O_2$ and $O_3$
which is a $1/m_b$-suppressed operator up to corrections of
$\mathcal{O}(\alpha_s^2)$ times a leading order operator.
\begin{equation}
\label{eq:R0def}
R_0 = O_2 + O_3 + \frac{O_1}{2} + \alpha_s (0.345 \,O_1 + 0.637 \, O_3) ,
\end{equation}
evaluating the radiative corrections at $\mu=m_b$.
We can obtain the matrix elements for this operator
through $\mathcal{O}(\alpha_s)$
from our lattice calculation. $R_1$ and $\tilde{R}_1$ are proportional
to $\mathcal{O}_4$ and $\mathcal{O}_5$ respectively
and will be discussed further below.

The matrix elements for $R_0$ in Eq.~(\ref{eq:R0def}) can be rewritten
in terms of our bag parameters using the definition in Eq.~(\ref{eq:etadef}):
\begin{eqnarray}
\label{eq:R0me}
\langle B_q | R_0 | B_q \rangle &=& -f_{B_q}^2M_{B_q}^2\left(\frac{M_{B_q}}{m_b(\mu)+m_q(\mu)}\right)^2 \times \\
&&\hspace{-3.5em}\left[\frac{5}{3}B^{(2)}_{B_q}(\mu) -\frac{1}{3}B^{(3)}_{B_q}(\mu)(1+0.637\alpha_s)-\right. \nonumber \\
&&\hspace{-2.0em}\left.\frac{4}{3}B^{(1)}_{B_q}(\mu)(1+0.690\alpha_s)\left(\frac{m_b(\mu)+m_q(\mu)}{M_{B_q}}\right)^2 \right] .\nonumber
\end{eqnarray}
Writing it in this way makes clear (setting $B^{(n)}_{B_q}$ to 1 and
$\alpha_s$ to zero) the expected cancellation at leading
order to leave matrix elements that are $\mathcal{O}(1/m_b)$.
Our evaluation of the term in square brackets above yields
\begin{eqnarray}
\label{eq:R0res}
\langle B_d | R_0 | B_d \rangle &=& -f_{B_d}^2M_{B_d}^2(3\eta^d_3) \times \rzerobagd \\
\langle B_s | R_0 | B_s \rangle &=& -f_{B_s}^2M_{B_s}^2(3\eta^s_3) \times \rzerobags \, .\nonumber
\end{eqnarray}
Note that the uncertainties here include both those for missing $\alpha_s^2$
terms in the matching of lattice QCD operators to continuum operators and
also the effect of missing $\alpha_s^2$ terms in the definition of $R_0$.
This latter uncertainty is estimated by calculating the size of the
$\alpha_s$ corrections in Eq.~(\ref{eq:R0def}) and multiplying by $\alpha_s$.
This gives a 35\% uncertainty, which dominates the error quoted in Eq.~(\ref{eq:R0res}).
To assist with numerical evaluation we have replaced the square ratio
of masses in Eq.~(\ref{eq:R0me}) with $3\eta^q_3$; values for this can be
found in Table~\ref{tab:eta_iq}. Eq.~(\ref{eq:R0res}) avoids use
of a perhaps somewhat arbitray definition of a bag parameter for $R_0$ given
in~\cite{Lenz:2006hd}. The numerical factors show clearly that this is a
$1/m_b$-suppressed operator by being of
size $\Lambda_{\text{QCD}}/m_b \approx$ 10\%.

$R_1$ and $\tilde{R}_1$ are defined as~\cite{Lenz:2006hd}
\begin{eqnarray}
\label{eq:r1def}
R^q_1 &=& \frac{m_q}{m_b} O_4 \\
\tilde{R}^q_1 &=& \frac{m_q}{m_b} O_5 \, .\nonumber
\end{eqnarray}
The matrix elements for $B_s$ and $B_d$
can then be determined from our results in Table~\ref{tab:OnMSB}.
Our bag parameters for $O_4$ and $O_5$ are given in
Table~\ref{tab:bagparam}. In~\cite{Lenz:2006hd} bag parameters
for $R_1$ and $\tilde{R}_1$ are defined in such a way as to set the
squared mass ratios in Eq.~(\ref{eq:etadef}) to 1. This means that the
bag parameters for $R_1$ and $\tilde{R}_1$ for the definition
in~\cite{Lenz:2006hd} can be recovered from our bag parameters by
multiplying by $3\eta_4^q/7$ and $3\eta_5^q/5$ respectively.
These factors are larger than 1.
Note however that the impact of both $R_1$ and $\tilde{R}_1$ on $\Delta \Gamma$
is tiny because of the $m_q/m_b$ factors in their definition.

Matrix elements of the numerically more important
$R_2$ and $\tilde{R}_2$ operators~\cite{Lenz:2006hd},
along with those of $R_3$ and $\tilde{R}_3$
cannot be directly obtained
from our current results because they contain derivatives
on the light quark fields inside the 4-quark operator.
Results of 
the calculations of these matrix elements are discussed 
separately in~\cite{Davies:2019gnp}.

\section{Conclusions}
\label{sec:conclusions}

We give results from the first `second-generation' lattice
QCD calculation of the matrix elements that contribute to
$B_s$ and $B_d$ mixing in and beyond the Standard Model.
We include $c$ quarks in the sea for the first time and have
a range of $u/d$ quark masses (taken to be equal) that
go down to the physical value. We use radiatively-improved NRQCD
for the $b$ quark action. By calculating the ratio of
the matrix elements of the 4-quark operators to the square of the decay
constant times mass (proportional to a quantity known as the bag
parameter) we obtain results with very little dependence on
the $u/d$ quark mass or the lattice spacing.
This gives us more accurate results than previous calculations for
these ratios, and the associated bag parameters, for
all five $\Delta B = 2$ operators.

Our key results are given in Tables~\ref{tab:OnMSB} and~\ref{tab:bagparam}.
Table~\ref{tab:OnMSB} gives the
ratio of matrix elements to $(fM)^2$ with our final physical values given
in the last row. These are the numbers that should be used to reconstruct
the 4-quark operator matrix elements by multiplying by $(f_{B_q}M_{B_q})^2$.
Table~\ref{tab:bagparam} converts these ratios into bag parameters, defined
in Eq.~(\ref{eq:etadef}). These numbers can be compared to unity,
the result expected in the vacuum saturation approximation. Our error budget for the
bag parameters is given in Table~\ref{tab:errbudget}. We have uncertainties
of 4--7\% for the individual bag parameters. This uncertainty is dominated
by missing higher orders in the perturbative matching to the continuum
4-quark operators. The uncertainty is reduced to around 2\% in the
ratio of bag parameters for $B_s$ to $B_d$, since this renormalisation
cancels. The correlations between results for different operators are given
in Table~\ref{tab:corr-os} of Appendix~\ref{sec:corr}. Our $B_s$ to
$B_d$ ratios
are now accurate enough to see that they are above 1 for $O_1$ and $O_3$
and below 1 for $)_4$ and $O_5$.

Our results for the key $O_1$ bag parameters that appear in SM phenomenology
are (repeating Eq.~(\ref{eq:bhatres}))
\begin{eqnarray}
\label{eq:bhatres2}
\hat{B}^{(1)}_{B_s} &=&  \bhats \\
\hat{B}^{(1)}_{B_d} &=&  \bhatd \nonumber \\
\frac{\hat{B}^{(1)}_{B_s}}{\hat{B}^{(1)}_{B_d}} &=&  \BBsBBdratio \, , \nonumber
\end{eqnarray}
where we give the RGI bag parameter as defined in Eq.~(\ref{eq:rgiconv}).
Multiplying by decay constant values obtained
using HPQCD's approach to $b$-physics with HISQ quarks~\cite{McNeile:2011ng}
by the Fermilab/MILC collaboration~\cite{Bazavov:2017lyh},
we also obtain (repeating Eq.~(\ref{eq:fsqbhatres}))
\begin{eqnarray}
\label{eq:fsqbhatres2}
f_{B_s}\sqrt{\hat{B}^{(1)}_{B_s}} &=&  \fsqrtbhats \\
f_{B_d}\sqrt{\hat{B}^{(1)}_{B_d}} &=&  \fsqrtbhatd \nonumber \\
\xi &=& \xifinal \, .\nonumber
\end{eqnarray}

In Section~\ref{sec:discussion} we discuss the phenomenology from
our results. We obtain values for $\Delta M$ for $B_s$ and $B_d$
in Eq.~(\ref{eq:ourdeltaM}) to be compared to experiment.

Alternatively, and more usefully, we
can combine our results with experiment to obtain the CKM elements
$|V_{ts}|$ and $|V_{td}|$ and their ratio. These values are given in
Eq.~(\ref{eq:vresults}) and Figure~\ref{fig:VtdVts} shows the constraints
they give in the $V_{td}$-$V_{ts}$ plane. Our results are the most accurate
determinations of these CKM elements using lattice QCD and show
good consistency with determinations from tree-level processes assuming
CKM unitarity. This means that we see no signs of
new physics in neutral $B$-meson
oscillations at this improved level of accuracy.

We derive results, by taking a ratio to $\Delta M$,
 for the branching fractions for $B_s$ and $B_d$
to decay to $\mu^+\mu^-$, a key mode for new physics searches at LHC.
Our results are given in Eq.~(\ref{eq:ourbmumu}) and~(\ref{eq:usbmumurat}).
Figure~\ref{fig:BrBsBd} shows a comparison of our predictions
to those from the recent Fermilab/MILC calculation of decay
constants~\cite{Bazavov:2017lyh} along with the current experimental
picture. This is encouraging for future tests of
new physics contributions to these rare decay processes.

Finally, we give values in Eq.~\ref{eq:R0res} and below Eq.~(\ref{eq:r1def})
for the matrix elements for the
$R_0$, $R_1$ and $\tilde{R}_1$ operators that contribute to the
SM prediction for the width difference $\Delta \Gamma$.

To improve accuracy further in future requires
improving the matching of the lattice QCD 4-quark operators to those
in the continuum; this is the dominant source of uncertainty in our
bag parameters. Since lattice QCD perturbation theory is so hard
a renormalisation method that can be implemented within the lattice
calculation and then matched perturbatively to $\overline{\text{MS}}$
in the continuum could be preferable. A symmetric momentum-subtraction
scheme (RI-SMOM) has been found to work well for kaon mixing
calculations~\cite{Garron:2016mva} but
attention must be paid to removing nonperturbative artefacts in these schemes
if high accuracy is to be achieved~\cite{Lytle:2018evc}.
Such a method would need to be implemented with a relativistic quark
action on lattices with fine enough lattice spacing to allow
quark masses close to that of the $b$ for $am\,^{<}_{\sim} 1$.
This has been a very successful strategy for HISQ quarks for $B$-meson
decay constants~\cite{McNeile:2011ng, McNeile:2012qf, Bazavov:2017lyh},
but in that case the decay constants calculated with
HISQ do not need any renormalisation. Calculating 4-quark operator matrix
elements is much harder, but still feasible. The ETM work with twisted
mass quarks~\cite{Carrasco:2013zta} is encouraging for this programme
(although they used $n_f=2$ gluon fields and RI-MOM renormalisation)
as is the work by RBC/UKQCD
using domain-wall quarks~\cite{Boyle:2018knm}
(although they have only calculated $B_s$ to $B_d$ ratios so far).
It seems clear that in the next few years improvements in this
direction will be possible, pushing uncertainties on bag parameters
down to the $\sim$2\% level. This will allow $|V_{ts}|$ and
$|V_{td}|$ to be determined to 1\%.

%%%%%%%%%%%%%%%%%%%%%%%%%%%%%%%%%%%%%%%%%%%%%%%%%%%%%%%%%%%%%%%%%
\vspace{2mm}
\noindent{{\bf Acknowledgements}} We are grateful to the MILC collaboration for the use of their
gauge configurations and code.
The results described here were obtained using the Darwin Supercomputer
of the University of Cambridge High Performance
Computing Service as part of STFC's DiRAC facility.
This work was funded by STFC, the Royal Society, the Wolfson Foundation
and the US DOE and National Science Foundation.
We thank Chris Bouchard, Elvira G\'{a}miz and Alex Lenz for useful discussions.
One of the authors (GPL) is grateful to the
Department of Applied Mathematics and Theoretical Physics, Cambridge
University, for their hospitality during the two visits when much of
this analysis was done.

\appendix

\section{Fitting Protocols}
\label{sec:fitting-protocols}
\begin{figure}
\includegraphics[width=0.65\hsize]{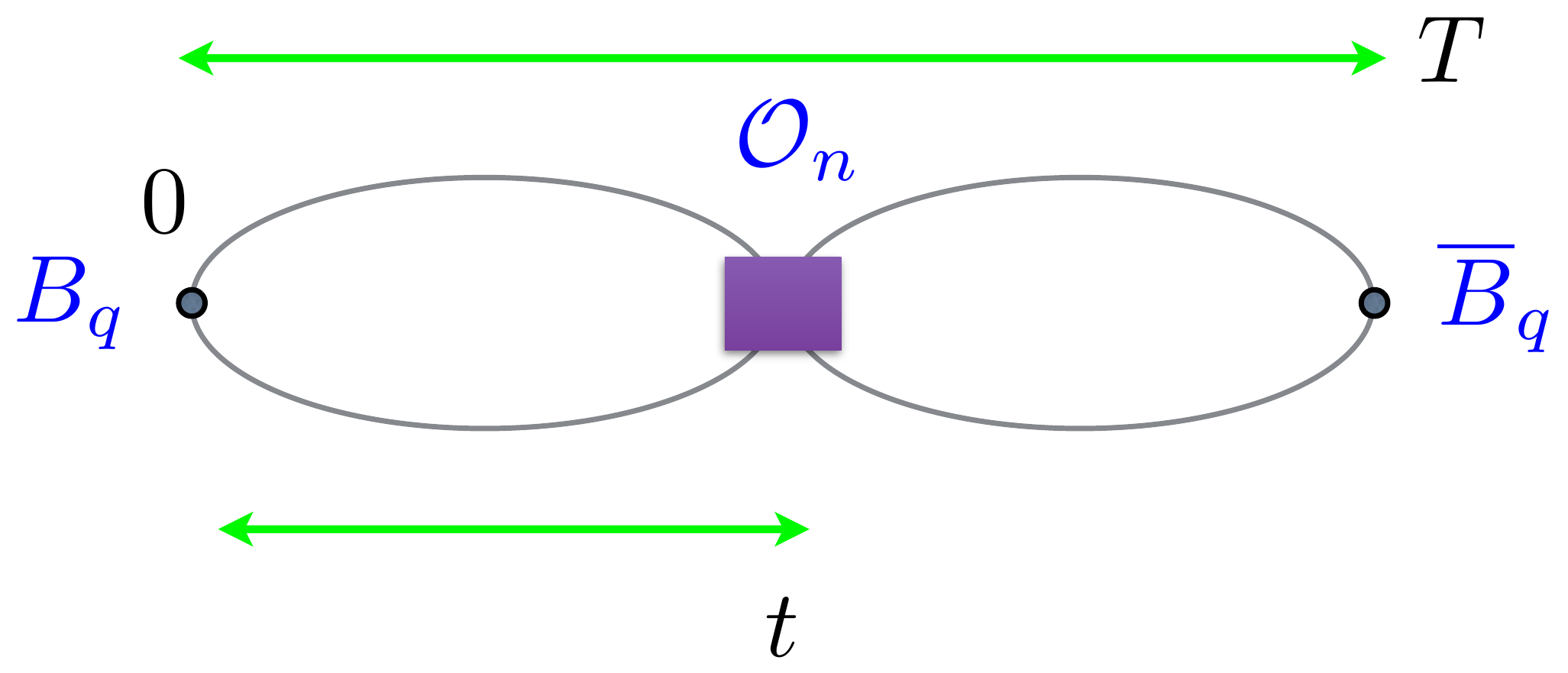}
\caption{\label{fig:3pt-pic} A schematic diagram of the 3-point function
for $B-\overline{B}$ mixing. $\mathcal{O}_n$ marks the insertion
at time $t$ of a 4-quark operator, when the meson and anti-meson
operators are located at times 0 and $T$.}
\end{figure}

We extract mixing amplitudes and decay constants by fitting Monte Carlo
data for 2-point and 3-point matrix correlators for each
meson:
\begin{align}
    \Gv(t) &\equiv \sum_{\xv} \langle 0 | \sv(\xv, t) \sv^T(0) | 0\rangle \\
    \Gv_\beta(t, T) &\equiv \sum_{\xv,\yv}
    \langle 0 | \sv(\xv, T) O_\beta(\yv, t) \sv^T(0)|0\rangle
\end{align}
where $\sv$ is a 3-vector of meson sources, the sums over
spatial~$\xv$ and~$\yv$ project onto zero 3-momentum, the
times satisfy~$0<t<T$, and $\beta=1, 2\ldots 5$ labels the mixing operator
(Figure~\ref{fig:3pt-pic}).
The sources include a local source,
corresponding to
\begin{equation}
    J_{A_0}^{(0)}=\overline\Psi_q\gamma_5\gamma_0\Psi_b^\nrqcd,
\end{equation}
and two smeared
sources: see~\cite{Dowdall:2013tga} for details. We also examine
the vector of correlators
\begin{equation}
        \Gv^{(1)}(t)
        \equiv \sum_{\xv} \langle 0 | J_{A_0}^{(1)}(\xv, t) \sv^T(0) | 0 \rangle
\end{equation}
where
\begin{equation}
    J_{A_0}^{(1)} \equiv
    -\frac{1}{2m_b}\Psi_q\gamma_5\gamma_0\boldsymbol{\gamma}\cdot\nabla\Psi_b^\nrqcd
\end{equation}
is the leading NRQCD correction to $J_{A_0}^{(0)}$.
We use the corrected current\footnote{Note that the
NRQCD-HISQ temporal axial current that we use here is correct through
the same order in $\alpha_s$ and $\Lambda_{\text{QCD}}/m_b$ as that of our 4-quark
operators. In~\cite{Dowdall:2013tga} we used a more
highly-corrected temporal axial current to determine $f_B$.}
to evaluate the decay constant (in lattice
units):
\begin{equation}\label{eq:A0matching}
    \big(1 + z_{A_0}\alpha_s\big)
    \langle 0|J^{(0)}_{A_0} + J^{(1)}_{A_0}|B_q\rangle = f_{B_q} M_{B_q}
\end{equation}
where coefficients $z_{A_0}$ depend on the lattice spacing and were
calculated in~\cite{Dowdall:2013tga} using~\cite{Monahan:2012dq}
(see Table~\ref{tab:zresults} for
the values we use here). The values for~$\alpha_s$
(from~\cite{Colquhoun:2015oha})
are given in Table~\ref{tab:params}.

The 2-point and 3-point correlators are calculated in the standard
way. For 3-point correlators this involves combining propagators
from a local source (at $t$) into `open-meson' propagators~\cite{Bazavov:2016nty}
which are then closed off at $0$ and $T$ (which cover all time-slices
away from $t$) with local or smeared meson
and anti-meson operators. We average correlators over 16 values of
$t$ for improved statistics. The smearing functions we use
are given in~\cite{Colquhoun:2015oha}.

Fits for the $B_s$ proceed in two steps. First the 2-point correlators
$\Gv(t)$ and $\Gv^{(1)}(t)$ are fit simultaneously. Then the best-fit
amplitudes and energies from that fit are used as priors for a
simultaneous fit of all the 3-point correlators~$\Gv_n(t)$.
Having finished the $B_s$ fits, we follow the same approach with the
$B_d$ correlators but constraining (via the priors) the $B_d$'s fit
parameters to be within 20\% of the corresponding values for the~$B_s$.
We discuss all of these fits in what follows.

\subsection{Fitting Two-Point Correlators}
We fit the 2-point correlators to a formula of the
form (in lattice units)
\begin{align}
    \Gv^\mathrm{fit}(t; \pv)
    &= \sum_{n=0}^{N-1} \Big( \mathrm{e}^{-E_n t} \cv_n \cv_n^T
    - (-1)^t \mathrm{e}^{-E_n^o t} \cv^o_n\cv_n^{oT} \Big), \\
    \Gv^{(1)\mathrm{fit}}(t)
    &= \sum_{n=0}^{N-1} \Big( \mathrm{e}^{-E_n t} j_n \cv_n^T
    - (-1)^t \mathrm{e}^{-E_n^o t} j^o_n\cv_n^{oT} \Big).
\end{align}
where the fit parameters~$\pv$ are comprised of all~$\cv_n$, $\cv_n^o$,
$E_n$, and~$E_n^o$.
Here~$\cv_n$ and~$\cv_n^o$ are 3-component vectors, and~$j_n$
and~$j_n^o$ scalars, where:
\begin{align}
\label{eq:cv}
    \cv_n &=  \frac{\langle 0 | \sv | E_n\rangle}{\sqrt{2M_n}}
    \quad&\quad
    \cv_n^o &= \frac{\langle 0 | \sv | E_n^o\rangle}{\sqrt{2M_n^o}},
    \\
\label{eq:jn}
    j_n &=  \frac{\langle 0 | J^{(1)}_{A_0} | E_n\rangle}{\sqrt{2M_n}}
    \quad&\quad
    j_n^o &=  \frac{\langle 0 | J^{(1)}_{A_0} | E_n^o\rangle}{\sqrt{2M_n^o}}.
\end{align}
In the exponents,
$E_n$ and $E_n^o$ are the energies of the lowest-lying states with
zero 3-momentum that couple to the sources.
The second (oscillating in time) term in
each correlator is due to taste-doubling caused by the staggered-quark
HISQ action for the light quarks
(see~\cite{Dowdall:2012ab}).
$M_n$ and $M_n^{o}$ are the physical masses corresponding to
states $|E_n\rangle$ and $|E_n^{o}\rangle$, respectively.
We
keep $N=6$ terms, but fit results are the same for any~$N\ge 5$.

We use a Bayesian fit procedure~\cite{gplbayes}.
Fits for the $B_s$~mesons on our coarsest lattices (0.15\,fm) use the
following Bayesian priors for the energies,
\begin{align}
    \log(E_0) &= \log(0.6(3)) & \log(\Delta E_n) &= \log(0.50(25)) \nonumber \\
    \log(E_0^o) &= \log(0.90(45)) & \log(\Delta E_n^o) &= \log(0.50(25)),
\end{align}
where $\Delta E_n\equiv E_n-E_{n-1}$, and
the logarithms indicate log-normal priors for energies.
Energies are rescaled in proportion to the lattice spacing to make
priors for the other lattices.
The priors for local and smeared amplitudes are
\begin{align}
    \log(\cv_n(\mathrm{loc.})) &= \log(0.2(8)) &
    \cv_n(\mathrm{smeared}) &= 1(4) \nonumber \\
    \log(\cv_n^o(\mathrm{loc.})) &= \log(0.2(8)) &
    \cv_n^o(\mathrm{smeared}) &= 1(4)
\end{align}
on the coarsest lattices. Amplitudes for the local source are rescaled
by~$a^{3/2}$ for the
other lattices. The smeared sources are designed to be lattice-spacing
independent and so we use the same prior for the other lattices.
Finally the priors for parameters $j_n$ and $j_n^o$ are
\begin{equation}
    j_n = -0.015(60) \quad\quad j_n^o = 0.02(8)
\end{equation}
on the coarse lattice, and again scale like~$a^{3/2}$ for other
lattices.

These central values for these priors were based
upon fit values for the ground
state~$B_s$. The uncertainties assigned to the priors
are large: for example, the priors are typically
2000--5000~times broader than the final fit errors for the
ground state parameters that we need for our analysis. Replacing
the central values by random values drawn from the prior distributions
leaves our results unchanged within errors.

We need to apply SVD cuts
to the data's correlation matrix because of the large number of correlators
being fit. The procedure for determining the SVD cuts is described in
Appendix~\ref{sec:SVD-cuts}; typically the cuts affect less than half of the data
modes. Fits for $\Gv(t)$ were constrained to $t$~values between the
values for $t_\mathrm{min}$ and $t_\mathrm{max}$
shown in Table~\ref{tab:fit-results}; data
from larger~$t$'s was too noisy to be useful. To keep the number of fit
points down (see Appendix~\ref{sec:SVD-cuts}), we restricted the fits
for $G^{(1)}(t)$ to the range of~$t$s between~$(t_\mathrm{min}+t_\mathrm{max})/2$
and~$t_\mathrm{max}$.

\begin{table*}
\caption{Simulation results
for $\langle {B}_q | O_n | \overline{B}_q \rangle_\latt /(f_{B_q} M_{B_q})^2$ for
$B_q=B_s,B_d$ mesons. Results are presented for each of the
configuration data sets described in Table~\ref{tab:gaugeparams}.
Fit ranges for 2-point ($t_\mathrm{min}\le t \le t_\mathrm{max}$)
and 3-point ($t_\mathrm{min}\le t \le T-t_\mathrm{min}$) correlators
are tabulated.
Sample $\chi^2$s per degree of freedom from fits to both sets
of correlators, with SVD and
prior noise (see Appendix~\ref{goodness-of-fit}), are also listed.
}\label{tab:fit-results}

\begin{ruledtabular}
\begin{tabular}{ccccccccccc}
meson & set & $t_\mathrm{min}$ & $t_\mathrm{max}$ & $T$ & $\langle{O}_1\rangle/(fM)^2$ & $\langle{O}_2\rangle/(fM)^2$ & $\langle{O}_3\rangle/(fM)^2$ & $\langle{O}_4\rangle/(fM)^2$ & $\langle{O}_5\rangle/(fM)^2$ & $\chi^2/\mathrm{dof}$ [dof] \\
\hline\hline
$B_s$ & 1 & 4 & 17 & 8--12  & $2.274(29)$ & $-1.887(24)$ & $0.3646(68)$ & $3.041(35)$ & $1.870(23)$& 0.97 [258] \\
$B_s$ & 2 & 4 & 17 & 8--12  & $2.336(29)$ & $-1.939(21)$ & $0.3625(75)$ & $3.199(33)$ & $1.978(22)$& 1.11 [258] \\
$B_s$ & 3 & 4 & 17 & 8--12  & $2.315(31)$ & $-1.923(26)$ & $0.3638(73)$ & $3.162(39)$ & $1.951(24)$& 1.01 [258] \\[1ex]
$B_s$ & 4 & 4 & 22 & 10--14, 17  & $2.367(27)$ & $-1.979(21)$ & $0.3720(70)$ & $3.382(34)$ & $2.060(22)$& 0.92 [324] \\
$B_s$ & 5 & 4 & 22 & 10--14, 17  & $2.319(26)$ & $-1.955(20)$ & $0.3636(70)$ & $3.288(34)$ & $1.994(22)$& 1.13 [324] \\
$B_s$ & 6 & 4 & 22 & 10--14, 17  & $2.365(32)$ & $-1.999(25)$ & $0.3747(80)$ & $3.366(40)$ & $2.037(25)$& 0.99 [324] \\[1ex]
$B_s$ & 7 & 5 & 33 & 12--16, 19  & $2.312(32)$ & $-2.029(29)$ & $0.3835(83)$ & $3.548(47)$ & $2.090(28)$& 1.03 [399] \\
\hline\hline
$B_d$ & 1 & 4 & 17 & 8--12  & $2.238(56)$ & $-1.869(39)$ & $0.362(14)$ & $3.054(59)$ & $1.861(41)$& 1.00 [258] \\
$B_d$ & 2 & 4 & 17 & 8--12  & $2.283(70)$ & $-1.875(43)$ & $0.345(19)$ & $3.232(69)$ & $2.016(49)$& 1.05 [258] \\
$B_d$ & 3 & 4 & 17 & 8--12  & $2.228(74)$ & $-1.854(49)$ & $0.324(19)$ & $3.232(77)$ & $1.955(53)$& 1.07 [258] \\[1ex]
$B_d$ & 4 & 4 & 22 & 10--14, 17  & $2.377(56)$ & $-1.941(38)$ & $0.333(14)$ & $3.539(67)$ & $2.158(44)$& 1.06 [324] \\
$B_d$ & 5 & 4 & 22 & 10--14, 17  & $2.332(59)$ & $-1.870(38)$ & $0.328(15)$ & $3.341(61)$ & $2.011(43)$& 0.99 [324] \\
$B_d$ & 6 & 4 & 22 & 10--14, 17  & $2.265(97)$ & $-1.965(67)$ & $0.379(25)$ & $3.49(10)$ & $2.086(71)$& 1.08 [324] \\[1ex]
$B_d$ & 7 & 5 & 33 & 12--16, 19  & $2.332(70)$ & $-1.984(68)$ & $0.356(18)$ & $3.68(10)$ & $2.153(63)$& 1.08 [399] \\
\end{tabular}
\end{ruledtabular}

\end{table*}
\subsection{Fitting Three-Point Correlators}
The fit function for the 3-point correlators is substantially more
complicated:
\begin{align}
    \Gv_\beta^\mathrm{fit}(t, T; \pv) &=
    \sum_{n,m=0}^{N-1} \mathrm{e}^{-E_nt}
    \cv_n \,V_{nm}(O_\beta)\,  \cv_m^T \mathrm{e}^{-E_m(T-t)}
    \nonumber \\
    &\hspace{-3.0em}- (-1)^{T-t}\sum_{n,m=0}^{N-1} \mathrm{e}^{-E_nt}
    \cv_n \,V_{nm}^o(O_\beta)\,\cv_m^{oT} \mathrm{e}^{-E_m^o(T-t)}
    \nonumber \\
    &\hspace{-3.0em}- (-1)^t\sum_{n,m=0}^{N-1} \mathrm{e}^{-E_n^ot}
    \cv_n^o \,V_{mn}^o(O_\beta)\,  \cv_m^T \mathrm{e}^{-E_m(T-t)}
    \nonumber \\
    &\hspace{-3.0em}+ (-1)^T\sum_{n,m=0}^{N-1} \mathrm{e}^{-E_n^ot}
    \cv_n^o \,V_{nm}^{oo}(O_\beta)\, \cv_m^{oT}
    \mathrm{e}^{-E_m^o(T-t)},
    \label{eq:Gv_beta}
\end{align}
where the fit parameters~$\pv$ include all of the 2-point correlator parameters
plus the~$V_{nm}$, $V_{nm}^o$, and~$V_{nm}^{oo}$.

We fit the 3-point amplitudes over the range
 $t_\mathrm{min}\le t \le T-t_\mathrm{min}$
for the values of~$T$ shown in Table~\ref{tab:fit-results}.
Parameters $\cv_n$, $\cv_n^o$, $E_n$, and~$E_n^o$ are the same as in the
2-point correlators; we use the results from the fits to the
2-point correlators
as priors for these parameters in our 3-point fits. The priors
on the coarsest lattices for each of
the mixing amplitudes $V(\beta)$, $V^o(\beta)$ and $V^{oo}(\beta)$
are:
\begin{align}
    V(O_1) &= 0.03(12) \quad&\quad V(O_2) &=  -0.03(12) \nonumber \\
    V(O_3) &= 0.005(20) \quad&\quad V(O_4) &=  0.04(16) \nonumber \\
    V(O_5) &= 0.025(100);
\end{align}
these are scaled in proportion to~$a^3$ for the other lattices.
Note that $V_{n,m}(\beta)$ and $V^{oo}_{n,m}(\beta)$ are symmetric
under interchange of~$n$ and~$m$. We are interested in the ground-state
value for
\begin{equation}
\label{eq:v00}
    V_{00}(O_\beta) = \frac{\langle E_0 | O_\beta | E_0 \rangle}{2M_0}.
\end{equation}

We introduce two simplifications to the analysis that make our fits run
20--100~times faster, without affecting fit results or precision.
The first simplification
is to replace both the data and the fit function in the fits with
their sums over~$t$,
\begin{align}
    \Gv_\beta(t,T) &\to  \sum_{t=t_\mathrm{min}}^{T-t_\mathrm{min}} \Gv_\beta(t,T)
    \nonumber \\
    \Gv_\beta^\mathrm{fit}(t,T;\pv) &\to
    \sum_{t=t_\mathrm{min}}^{T-t_\mathrm{min}} \Gv_\beta^\mathrm{fit}(t,T;\pv),
\end{align}
while keeping the same fit parameters~\cite{bouchard}.
This reduces the number of data points
to be fit for our~0.09\,fm lattice, for example,
from~1050 to~180.
Note that the Monte Carlo
data for~$\Gv_n(t,T)$ do not vary
much with~$t$, as expected from~\Eq{Gv_beta}.

The second simplification is to marginalize all fit parameters other
than those associated with the ground state~\cite{Hornbostel:2011hu}. We
do this by splitting the fit function (\Eq{Gv_beta}) into two parts, one
that involves only the ground state (i.e., either the $B_s$ or $B_d$) and
the other with the remaining terms:
\begin{equation}
    \Gv_\beta^\mathrm{fit} \equiv
    \cv_0 \,V_{00}(O_\beta)\,  \cv_0^T \mathrm{e}^{-E_0T}
    + \Delta\Gv_\beta^\mathrm{fit} \, .
\end{equation}
We then replace the fit data~$\Gv^\mathrm{lat}_\beta$ by
\begin{equation}
    \Gv^\mathrm{lat}_\beta(t, T) \to \Gv^\mathrm{lat}_\beta(t, T)
    -\Delta \Gv_\beta^\mathrm{fit}(t,T;\pv_\mathrm{prior}),
\end{equation}
where the prior values for the fit parameters are used
in~$\Delta \Gv_\beta^\mathrm{fit}$.
At the same time, we replace the fit function by just its ground-state
term:
\begin{equation}
    \Gv_\beta^\mathrm{fit}(t,T;\pv) \to
    \cv_0 \,V_{00}(O_\beta)\,  \cv_0^T \mathrm{e}^{-E_0T} \, .
\end{equation}
This reduces the number of fit parameters from~450 to~9 (for $N=6$).
Marginalization works particularly well here because we have excellent
priors for the amplitudes and energies, from the 2-point correlators,
and because the mixing parameters enter the fit function linearly.
Also the marginalized fit function is $t$-independent, making the
first simplification (summing over~$t$) quite natural.

Again we need SVD cuts, but the need is greatly reduced by
summing over~$t$. We used the method outlined in Appendix~\ref{sec:SVD-cuts};
typically the cuts modified around 70\% of the data
modes.

We tabulate simulation results (using Eqs.~(\ref{eq:A0matching}),~(\ref{eq:cv}),~(\ref{eq:jn}) 
and~(\ref{eq:v00})) for the dimensionless ratio
\begin{equation}
    \frac{\langle B_q | O_n | B_q \rangle}{f_{B_q}^2 M_{B_q}^2}
\end{equation}
with $B_q=B_s,B_d$ in Table~\ref{tab:fit-results}.
This table also shows sample values of $\chi^2$ from the various (2-point and 3-point)
correlator fits when we include random SVD and prior noise, as
discussed in Appendix~\ref{goodness-of-fit}. Without noise,
the $\chi^2$s per degree
of freedom are much smaller than~1.0, as expected.
Also following Appendix~\ref{goodness-of-fit}, we tested the
uncertainties from our fits using simulated data. Fits to simulated
data reproduced the input parameters to within errors.

We verified that marginalization and averaging over~$t$ have negligible
effect on our results. With our smallest lattice spacing (0.09\,fm),
for example, undoing both optimizations gives the following values:
\begin{equation}
     \frac{\langle B_s | O_n | B_s \rangle}{f_{B_s}^2 M_{B_s}^2}
     = \begin{cases}
     2.307(35) & n=1\\
     -2.025(31) & n=2 \\
     0.3847(86) & n=3 \\
     3.547(51) & n=4 \\
     2.094(31) & n=5 \, .
     \end{cases}
 \end{equation}
These agree well with the values in Table~\ref{tab:fit-results} (for set~7),
but took far longer to compute.

\begin{table}
\caption{Sample error budgets from simulations on the 0.09\,fm lattice
(set~7) for
$\langle B_q | O_n | B_q \rangle /(f_{B_q} M_{B_q})^2$ for
$B_s$ and $B_d$ mesons. Percentage errors coming from Monte Carlo statistics,
the fit priors, and the SVD cuts are shown; these are added in quadrature to
give the total error.}
\label{tab:On-error-budget}
\begin{ruledtabular}
\begin{tabular}{cccc}
$B_s$ & $\langle{O}_1\rangle/(fM)^2$ & $\langle{O}_2\rangle/(fM)^2$ & $\langle{O}_3\rangle/(fM)^2$ %& $\langle{O}_4\rangle/(fM)^2$ & $\langle{O}_5\rangle/(fM)^2$
\\ \hline
    statistics &     1.25 &   1.15   &  2.00  %&    1.15   &   1.16
     \\
    prior &     0.28 &   0.31   &  0.47
    \\
      SVD &     0.42 &   0.74   &  0.56
      \\ \hline
    total    &  1.35   &   1.41    &  2.13
\end{tabular}
\end{ruledtabular}
\\[2ex]
\begin{ruledtabular}
\begin{tabular}{cccc}
$B_d$ & $\langle{O}_1\rangle/(fM)^2$ & $\langle{O}_2\rangle/(fM)^2$ & $\langle{O}_3\rangle/(fM)^2$ %& $\langle{O}_4\rangle/(fM)^2$ & $\langle{O}_5\rangle/(fM)^2$
\\ \hline
     statistics   &  2.59  &    2.13  &    4.25  \\
    prior   &  0.89  &    1.00  &    1.87  \\
      SVD   &  1.38  &    2.53  &    1.75  \\
      \hline
    total &      3.07   &   3.46   &   4.96
\end{tabular}
\end{ruledtabular}
\end{table}

In Table~\ref{tab:On-error-budget}, we show sample error budgets
for these quantities from simulations on the 0.09\,fm~lattice;
others are similar. The dominant source of uncertainty is from the
Monte Carlo statistics.

\subsection{Blind Analysis}
This analysis was blinded by multiplying the 3-point correlators by
a random normalization factor. The random factor was removed only
after the entire analysis was completed and this paper written.

%%%%%%%%%%%%%%%%%%%%%%%%%%%%%%%%%%%%%%%%%%%%%%%%%%%%%%%%%%%%%%%%%
%
\section{Chiral Fit}
\label{sec:chifits}
%
%%%%%%%%%%%%%%%%%%%%%%%%%%%%%%%%%%%%%%%%%%%%%%%%%%%%%%%%%%%%%%%%%

Although we have results at physical pion masses we do not
rely on these simply for our final value. We include also results
at heavier-than-physical pion masses, which are statistically more precise,
by using a fit to the dependence on the pion mass based
on chiral perturbation theory. This gives the coefficients of
the non-analytic `chiral logarithms' in $m_{\pi}^2\log(m_{\pi}^2)$;
in addition we include analytic
terms to allow both for staggered quark discretisation effects, the unphysically
heavy $u/d$ quark masses in the sea and for mistuning of valence quark masses.
Performing a fit to results at multiple pion masses then tests
the dependence expected from chiral perturbation theory.

Our principal results consist of values for
the ``reduced'' matrix elements of the 4-quark operators,
\begin{equation} \label{eq:red-mat-element}
    R_{q}^{n} \equiv
    \frac{\langle {B}_q | O_n | \overline{B}_q \rangle_\msb^{(m_b)}}{\big(f_{B_q} M_{B_q}\big)^2},
\end{equation}
on each configuration set (Table~\ref{tab:OnMSB}).
We fit the $R^n_q$
to the form
\begin{eqnarray}
\label{eq:chifit}
R(m_l,m_s) &=& R(m_l^{\text{phys}},m_s^{\text{phys}})\times \\
&&\hspace{-5.0em}\left( 1 + p_{\text{log}}\chi_{\text{log}} + p_J 3g^2 J +
p_{a^4}\delta X_{a^4} + p_{m_\pi^2a^2} \delta X_{m_{\pi}^2a^2} \right. \nonumber \\
&&\left. p_l\delta x_l + p_{l2}(\delta x_l)^2 + p_v \delta_v   \right) \nonumber
\end{eqnarray}
where we suppress the indices $n$ and $q$ on each term and each
parameter, $p$, for clarity. The functions used in each term are discussed
below. $\chi_{\text{log}}$ is given in Eq.~(\ref{eq:chilog}); $J$ in
Eq.~(\ref{eq:chiJ}); $\delta X$ in Eq.~(\ref{eq:deltaX}); $\delta x_l$ in
Eq.~(\ref{eq:deltaxl}) and $\delta_v$ in Eq.~(\ref{eq:deltav}).
Note that this fit is done after applying the additional uncertainties discussed
in Section~\ref{sec:simerr} to allow for matching and discretisation effects.

To derive this form, we make use of the
results in~\cite{Bernard:2013dfa}, where Appendix A gives the
dependence on light meson masses of the bag parameters at one-loop
in heavy meson staggered chiral perturbation theory.
This builds on the continuum heavy meson chiral perturbation theory
results of~\cite{Detmold:2006gh}.
There is a lot of cancellation of chiral logarithms between
4-quark operator matrix elements and decay constants so that, as we
discuss below, the remaining chiral logarithm terms ($\chi_{\text{log}}$ and
$J$ in Eq.~\ref{eq:chifit}) in the bag parameters (and equivalently in $R$)
have small coefficients. This expected very benign dependence on the
light quark mass is another reason for working with the bag parameters
as we do here, rather than the 4-quark operator matrix elements.

\begin{figure}
\includegraphics[width=0.85\hsize]{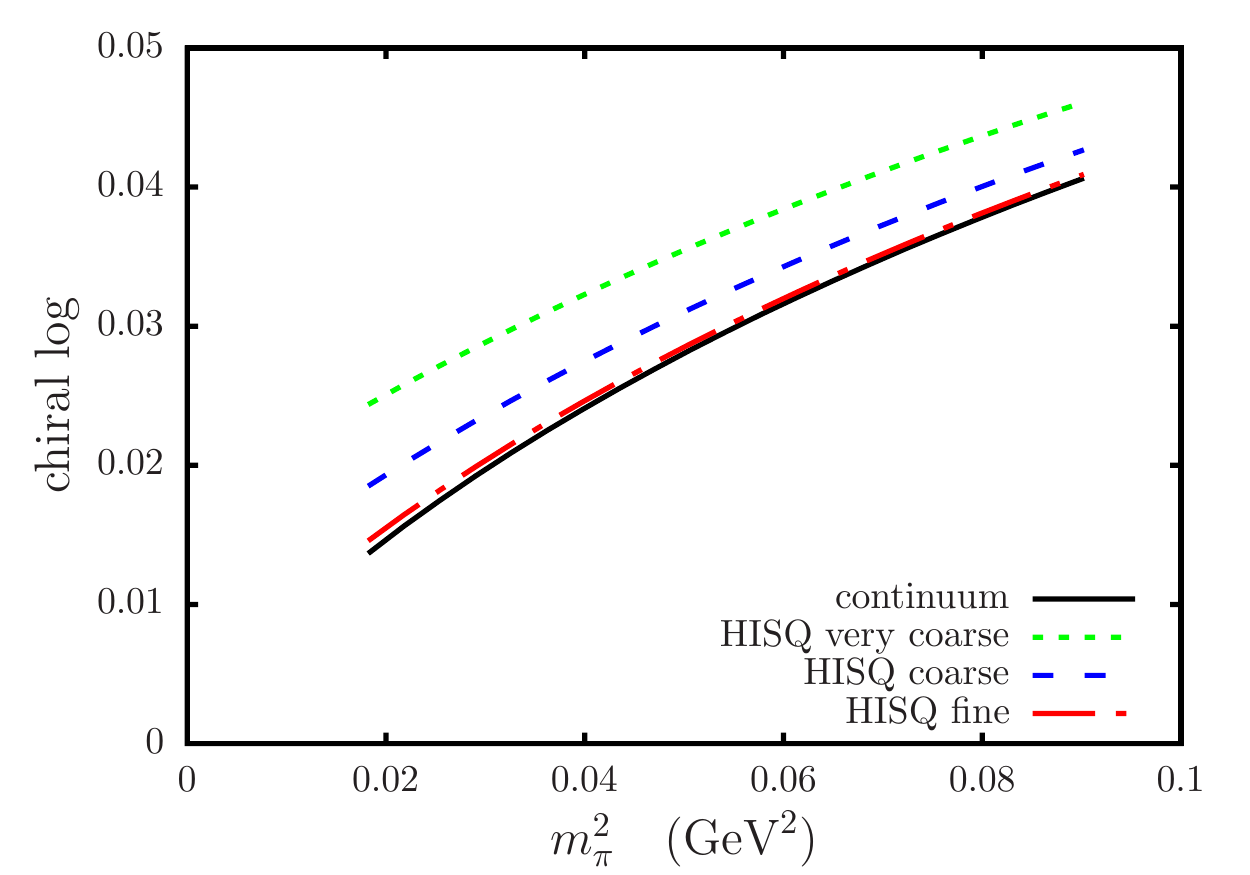}
\caption{\label{fig:chiral}
A comparison of chiral logarithm terms
in $m_{\pi}$ that appear in the continuum chiral perturbation theory
for $R_d$
(Eq.~(\ref{eq:chitadcont}))
with those in
staggered chiral perturbation theory (Eq.~(\ref{eq:chitad})) for HISQ quarks.
The solid black line gives the continuum form (Eq.~\ref{eq:chitadcont})
and the dashed blue and red curves give the staggered form
in Eq.~(\ref{eq:chitad}) for HISQ quarks on very coarse, coarse and fine lattices,
respectively.
}
\end{figure}

The chiral perturbation theory for the bag parameters is
given in~\cite{Bernard:2013dfa} in the form (using our
notation)
\begin{equation}
\label{eq:claude}
R^n_q = \beta_n \left(1 \pm S^q+\tilde{T}^q_n + \frac{\beta_n^{\prime}}{\beta_n}(Q^q_n + \tilde{Q}^q_n)\right) \, .
\end{equation}
Here $\beta_n$ is the low-energy constant (value at
zero pion mass) for $R^n$ and $\beta_n^{\prime}$ is the
equivalent term for 4-quark matrix elements between vector
heavy-light mesons. For $O_1$, $\beta_1^{\prime}=\beta_1$.
$S$ comes from `tadpole' diagrams (with $+$ for $n$ = 1, 2 and 3
and $-$ for $n$ = 4, 5) and
$Q$ from `sunset' diagrams that connect pseudoscalar and
vector mesons. $\tilde{T}$ and $\tilde{Q}$ are `wrong-spin'
tadpole and sunset terms respectively.
Below we discuss the content of these functions in terms
of the important non-analytic chiral logarithms and the
effect on these of the discretisation effects in the staggered quark
formalism. This enables us to transcribe Eq.~(\ref{eq:claude})
into the simpler Eq.~(\ref{eq:chifit}) that we will use.
We now discuss each of these terms in turn.

\subsection{Tadpole diagrams}
\label{subsec:tadpoles}

The results in~\cite{Bernard:2013dfa} are given in terms
of meson masses that include those for pions of different taste that appear in
the staggered quark formalism. Thus a term that would be
a simple chiral logarithm in the continuum can appear in a number of guises,
one of which is as an
average over the masses
of all tastes of pion. On fine enough lattices this will become a continuum
logarithm plus discretisation effects. In fact, for the fully unquenched
case that we study here ($m^{\text{val}}=m^{\text{sea}}$), staggered chiral
perturbation theory typically arranges itself to cancel taste effects inside
chiral logarithms so that non-analyticities in $a^2$ cancel
as $m_{u/d} \rightarrow 0$ (see Appendix A of~\cite{Colquhoun:2015mfa}).
This also happens here.
The chiral logarithm in $R_d$ from tadpole diagrams ($S$ in Eq.~(\ref{eq:claude}))
appears in the form
\begin{equation}
\label{eq:chitad}
-\frac{1}{16}\sum_{\text{tastes,t}} \frac{m^2_{\pi,t}}{\Lambda^2_{\chi}}\log{\frac{m^2_{\pi,t}}{\mu_{\chi}^2}} +
\frac{1}{2} \frac{m^2_{\pi,I}}{\Lambda^2_{\chi}}\log{\frac{m^2_{\pi,I}}{\mu_{\chi}^2}}.
\end{equation}
Here $I$ denotes the singlet (largest mass) pion taste. We can compare
this function to the corresponding continuum chiral logarithm
\begin{equation}
\label{eq:chitadcont}
-\frac{1}{2} \frac{m^2_{\pi,P}}{\Lambda^2_{\chi}}\log{\frac{m^2_{\pi,P}}{\mu_{\chi}^2}} \, ,
\end{equation}
where $P$ denotes the lightest (Goldstone) pion taste. This comparison is shown
in Figure~\ref{fig:chiral} using taste-splittings for HISQ pions for the
lattice spacing values that we use in this calculation. We give
results for our range
of $m^2_{\pi} \equiv m^2_{\pi, P}$ values from the physical point, 0.018 $\mathrm{GeV}^2$, to 0.09 $\mathrm{GeV}^2$.
$\Lambda_{\chi}=4\pi f_{\pi}$ = 1.64 GeV and we take $\mu_{\chi}=1.0$ GeV.
The difference between the HISQ and continuum
chiral logarithm terms is sufficiently small that we simply allow for that
discrepancy in our treatment of discretisation effects. This is included
through the $\delta X$ terms in Eq~(\ref{eq:chifit}) discussed below.

We therefore take the chiral logarithm terms in Eq.~(\ref{eq:chifit})
with continuum form:
\begin{eqnarray}
\label{eq:chilog}
\chi^d_{\text{log},n} &=& -\frac{1}{2}\frac{m_{\pi}^2}{\Lambda^2_{\chi}}\log\left(\frac{m_{\pi}^2}{\mu_{\chi}^2}\right) - \frac{1}{6}\frac{m_{\eta}^2}{\Lambda^2_{\chi}}\log\left(\frac{m_{\eta}^2}{\mu_{\chi}^2}\right) - (\mathrm{phys}) \nonumber \\
\chi^s_{\text{log},n} &=& -\frac{2}{3}\frac{m_{\eta}^2}{\Lambda^2_{\chi}}\log\left(\frac{m_{\eta}^2}{\mu_{\chi}^2}\right) - (\mathrm{phys}) .
\end{eqnarray}
Here $m^2_{\eta}=(2m_{\eta_s}^2+m_{\pi}^2)/3$ and we use masses of Goldstone taste $\pi$
and $\eta_s$ in this expression.
$(\mathrm{phys})$ denotes the value of the previous expression evaluated
for physical masses so that the total right-hand side vanishes at that point.
$m^2_{\eta}$ changes very little as $m_{\pi}$ changes so these
terms in the fit do very little.
The parameters $p^q_{\text{log},n}$ are given priors $+1.0(3)$ for $n=1,2,3$
and $-1.0(3)$ for $n=4,5$ as this logarithm appears with opposite sign
for $\mathcal{O}_{4,5}$. The prior width allows for modification of the coefficients
from missing higher order terms in chiral perturbation theory.

\begin{figure}
\includegraphics[width=0.85\hsize]{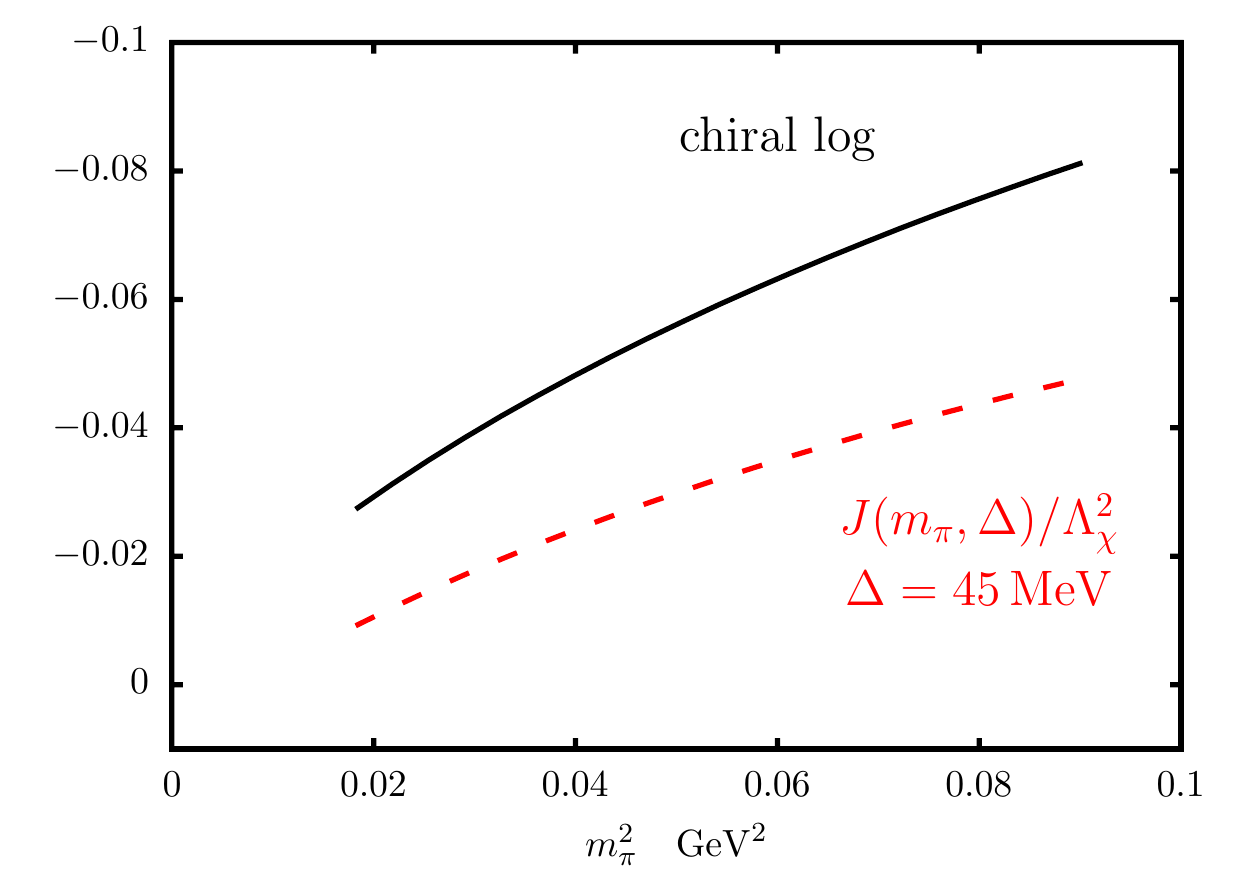}
\caption{\label{fig:jhyp} A comparison of the function
$J(m_{\pi},\Delta)/\Lambda^2_{\chi}$ to the chiral logarithm
to which it is equal at $\Delta=0$
($(m_{\pi}^2/\Lambda_{\chi}^2)\log(m_{\pi}^2/\mu_{\chi}^2)$) as a function of $m_{\pi}$.
}
\end{figure}

\subsection{Sunset diagrams}
\label{subsec:sunset}

We now turn to the term denoted $J$ in Eq.~(\ref{eq:chifit}) that comes
from the sunset diagram term $Q$ in Eq.~(\ref{eq:claude}).
This would also take the form of a chiral logarithm
$m_{\pi}^2\log(m^2_{\pi}/\mu_{\chi^2}^2)$ in the
infinite heavy meson mass limit in the continuum.
For finite heavy meson mass, however, $J$ is modified by terms that depend on
heavy meson mass differences,
because there is a pseudoscalar to
vector heavy meson transition inside the sunset diagram. The form of $J$ as a function
of pion mass and heavy meson mass difference, $\Delta$,
is given in Eq.~6.17 of~\cite{Bazavov:2011aa},
which considers chiral perturbation theory terms for
the heavy-light meson decay constant.
In that case the appropriate value for $\Delta$ can
include heavy-strange to heavy-light mass differences as well as vector to pseudoscalar
mass differences. Here, when we consider $R_d$ and $R_s$, that does not happen and
we only have to consider the case where
$\Delta = M_{B_{(s)}^*}-M_{B_{(s)}}$. Then $\Delta$ takes the value 45 MeV for
$B^*-B$~\cite{Tanabashi:2018oca} and we take the same value for $B^*_s-B_s$ since
any differences are expected~\cite{Dowdall:2012ab} and seen to
be~\cite{Tanabashi:2018oca} small. Figure~\ref{fig:jhyp} compares the function
$J(m_{\pi},\Delta)/\Lambda^2_{\chi}$ with that of the chiral
logarithm to which it is equal
when $\Delta=0$. Even though $\Delta$ is small, and much
smaller than $m_{\pi}$ through most of the
range in which we work,
we see $\Delta$ does have an impact, so that $J$ has smaller
magnitude and gradient in $m^2_{\pi}$
than its associated chiral logarithm.

$J$ in Eq.~(\ref{eq:chifit}) then takes the form
\begin{eqnarray}
\label{eq:chiJ}
J^d_{n} &=& \frac{1}{2}\frac{J(m_{\pi},\Delta)}{\Lambda^2_{\chi}}+ \frac{1}{6}\frac{J(m_{\eta},\Delta)}{\Lambda^2_{\chi}} - (\mathrm{phys}) \nonumber \\
J^s_{n} &=& \frac{2}{3}\frac{J(m_{\eta},\Delta)}{\Lambda^2_{\chi}}- (\mathrm{phys})\, .
\end{eqnarray}
where $J(m,\Delta)$ is given in~\cite{Bazavov:2011aa}. $J$ is multiplied by
3$g^2$, where $g$ is the $BB^*\pi$ coupling. We take the value
of $g$ as 0.5, based on recent lattice QCD
calculations~\cite{Detmold:2011bp,Bernardoni:2014kla,Flynn:2015xna}.
The uncertainty on $g$, both from the lattice calculations but also from the effect of
missing higher order terms in chiral perturbation theory, is absorbed into
the coefficient $p_J$. $p_J$ is the ratio of low-energy constants associated with
the bag parameters for vector heavy-light mesons to that for pseudoscalars.
For $\mathcal{O}_1$ we know that this ratio is 1~\cite{Grinstein:1992qt}. For the
other operators, $n$=2--5, we do not. We therefore take a prior value and width
on $p^{(d,s)}_{J,n}$ of 0(1), allowing either sign.
For $p^{(d,s)}_{J,1}$ we take 1.0(3) to allow for
uncertainty in $g^2$. Note that in the case where $\Delta=0$ the coefficient
of the chiral logarithm $(m^2_{\pi}/\Lambda^2_{\chi}\log(m^2_{\pi}/\mu^2_{\chi})$
in the chiral perturbation theory for $O_1$
is $(3g^2-1)/2$~\cite{Becirevic:2006me}, which is small (-0.125) when $g=0.5$.

\begin{figure}
\includegraphics[width=0.85\hsize]{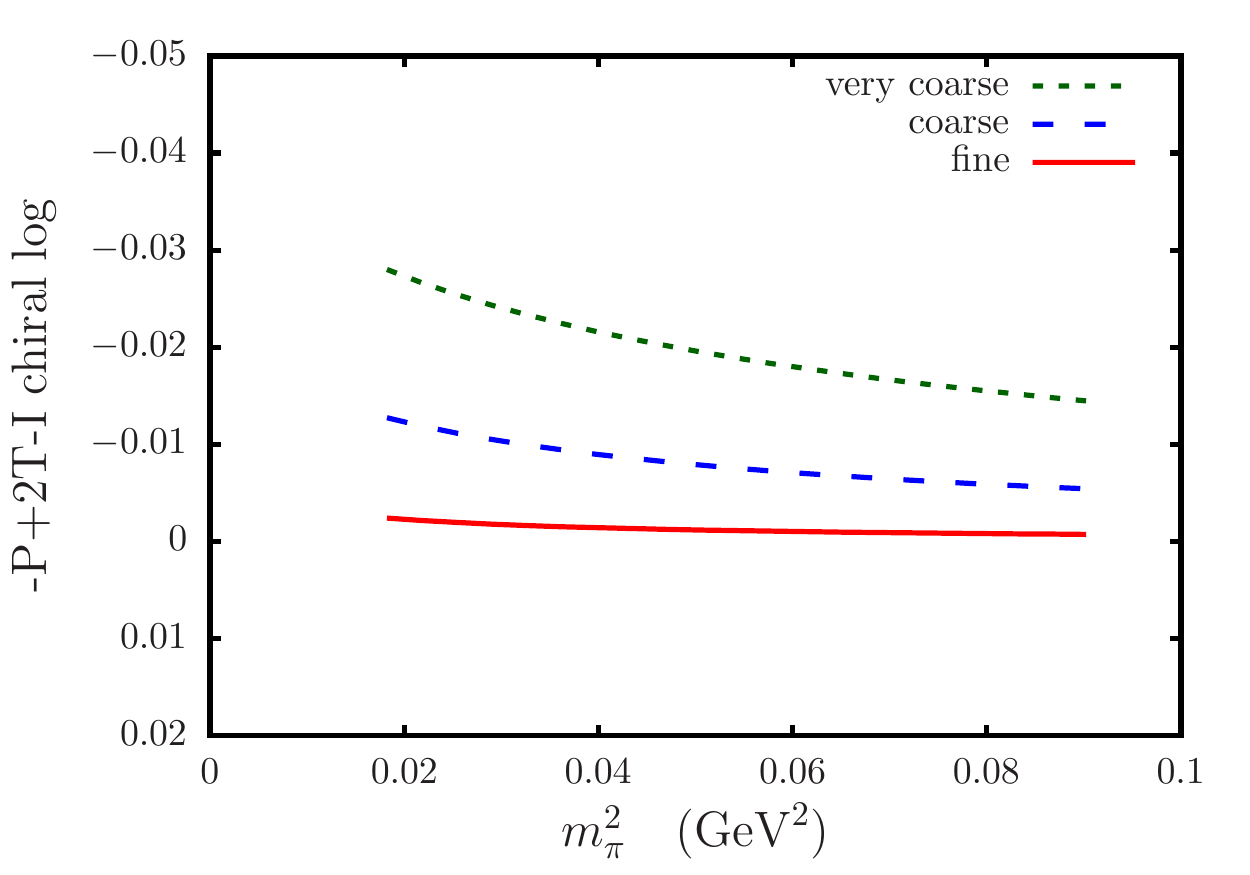}
\caption{\label{fig:pti-tilde} The function $\tilde{X}$, defined
in Eq.~(\ref{eq:ptitilde}), plotted for HISQ light quarks
against the square of the
pion mass. We give three curves,
for lattice spacing values corresponding to our
very coarse, coarse and fine lattices.
This function appears in the `wrong-sign' tadpole terms in
heavy meson staggered
chiral perturbation theory.
}
\end{figure}

\subsection{Wrong-spin and other effects}
\label{subsec:wrongspin}

The heavy meson staggered chiral perturbation theory analysis
of~\cite{Bernard:2013dfa} showed that $O_{1,2,3}$ and $O_{4,5}$ can
mix through `wrong-spin' staggered taste-effects ($\tilde{T}$ and
$\tilde{Q}$ in Eq.~(\ref{eq:claude})). The size of these terms
depends on the size of the light meson taste-splittings for the
staggered action. For the asqtad action used for light
quarks in~\cite{Bazavov:2016nty} they were of some concern and it
was important to include these effects explicitly in a full fit to
all 5 operators.
For the HISQ action that we use here these effects are much smaller.
The question then becomes whether they are distinct in magnitude or
form from discretisation effects from other sources that are already
included in our analysis.

The wrong-spin contributions from tadpole
diagrams for $B_d$ involve differences of chiral logarithms for different
taste pions (hence cancelling in the absence of taste-splitting
effects) along with hairpin
terms that have coefficients $a^2\delta_V^{\prime}$ and $a^2\delta_A^{\prime}$
that are themselves the size of the unit of
taste-splitting~\cite{Colquhoun:2015mfa}.
As an illustration of the impact of these terms we examine the
terms that are differences of chiral logarithms.
These appear in 3 `signatures' : $V-A$, $P-I$ and $-P+2T-I$. Here the
letter denotes the pion taste, ordered in increasing mass
as: $P$, $A$, $T$, $V$, $I$. Figure~\ref{fig:pti-tilde} illustrates the
size and behaviour of these terms for
the $-P+2T-I$ example that mixes $O_2$ and
$O_3$, the simplest because
there are no additional hairpin corrections.
The function plotted is
\begin{eqnarray}
\label{eq:ptitilde}
\tilde{X} &=& -\frac{m_{\pi,P}^2}{\Lambda^2_{\chi}}\log\left(\frac{m_{\pi,P}^2}{\mu_{\chi}^2}\right) + 2\frac{m_{\pi,T}^2}{\Lambda^2_{\chi}}\log\left(\frac{m_{\pi,T}^2}{\mu_{\chi}^2}\right) \nonumber \\
&-& \frac{m_{\pi,I}^2}{\Lambda^2_{\chi}}\log\left(\frac{m_{\pi,I}^2}{\mu_{\chi}^2}\right) \, .
\end{eqnarray}
We see that $\tilde{X}$ falls rapidly with lattice spacing (approximately as $(a\Lambda)^4$
and has a slope with $m^2_{\pi}$ that also falls with lattice spacing
(approximately as $(a\Lambda)^2$). This behaviour is generic for terms that
arise from taste-splittings in this way.

The wrong-spin tadpole terms have
a variety of coefficients multiplying them that correspond to ratios
of
4-quark operator matrix elements (within the groupings ${1,2,3}$ and ${4,5}$).
Most of the coefficients for the wrong-spin tadpole terms appearing
in the chiral expansion for $O_y$ are of
the form $\beta_x/(4\beta_y)$ where
$\beta_x$ is
the low energy constant associated with operator $O_x$.
The exception is $O_1$, where the coefficient is $2(\beta_2+\beta_3)/\beta_1$.
If all the 4-quark operator matrix elements were of the same size then the
coefficients would be $\mathcal{O}(1/4)$. $O_3$ and $O_5$ have smaller
matrix elements than the others, however, if
we consider the vacuum saturation approximation. Hence $\beta_x/\beta_y$ can
be of size 2 for $O_5$ and 5 for $O_3$.

We are already including $a^2$ and $a^4$ errors in \Eq{a2-errors},
but contributions from wrong-sign tadpole terms
differ between $B_d$ and $B_s$~mesons.
The largest contributions are for the $B_d$ meson; we allow for them
and similar terms that arise from sunset diagrams (and so contain differences
of $J(m_{\pi,t},\Delta)$) along with residual
right-sign taste-effects by including terms
\begin{eqnarray}
\label{eq:deltaX}
\delta X^{d}_{a^4} &=& \big(a\Lambda\big)^4  \\
\delta X^d_{m_\pi^2a^2} &=& \alpha_s \big(a\Lambda\big)^2 \delta_{m_{\pi}^2} \nonumber
\end{eqnarray}
in the chiral fit, Eq.~(\ref{eq:chifit}). We include them 
with the chiral fit rather than with other discretisation effects 
in~\Eq{a2-errors} because they arise from the staggered quark 
action and hence carry no $am_b$ dependence.  
Here
\begin{equation}
  \delta_{m_{\pi}^2} \equiv \frac{m_\pi^2 - 0.054}{0.072}
\end{equation}
allows for $m_\pi^2$~dependence; it
varies between $-0.5$ and~$0.5$ over our range of parameters.
Similar terms are not needed for $B_s$~mesons because the effects
are smaller and can be simulated by~\Eq{a2-errors}.
We take the priors
$p_{a^4}$
and $p_{m_{\pi}^2a^2}$ to be~0(2) since, although this is an
unnecessarily broad prior for some $n$, it allows a reasonable size
for all the possibilities.

\subsection{Analytic terms}
\label{subsec:analytic}

The three terms given symbol $\delta$ on the last line of Eq.~(\ref{eq:chifit})
are simple polynomials to account for
mistuning of sea and valence quark masses from their physical values.
We use
\begin{equation}
\label{eq:deltaxl}
\delta x_l = \frac{1}{10}\left(\frac{m_l}{m_s} - \left.\frac{m_l}{m_s}\right|_{\text{phys}}\right)
\end{equation}
where $m_l$ and $m_s$ are the sea $u/d$ and $s$ quark masses from
Table~\ref{tab:gaugeparams}.
The physical value for the $m_l/m_s$ ratio we take as 27.18(10) from~\cite{Bazavov:2017lyh}.
The factor of 1/10 converts $m_l/m_s$ to the size of the parameters
that appear in chiral perturbation theory as a ratio of meson masses
to $\Lambda_{\chi}=4\pi f_{\pi}$. Eq.~(\ref{eq:chifit}) includes terms
in $\delta x_l$ and $(\delta x_l)^2$. We take a prior of size 0(1)
for each coefficient $p_l$ and $p_{l2}$ (for each operator and
each $q$). We do not allow for mistuning effects for $c$ quarks in
the sea since we expect thse effects to be negligible compared to those
from light sea quarks.
$\delta_v$ accounts for the mistuning
of light and strange valence masses appropriate to $B_d$ or $B_s$. We take
\begin{eqnarray}
\label{eq:deltav}
\delta_v^l &=&   \frac{m^2_{\pi} - m^2_{\pi,\text{phys}}}{\Lambda^2_{\chi}}                   \\
\delta_v^s &=&   \frac{m^2_{\eta_s}-m^2_{\eta_s,\text{phys}}}{\Lambda^2_{\chi}} \, .                    \nonumber
\end{eqnarray}
Again the coefficients for this term, $p^q_{v,n}$, have prior 0(1) for
each $n$ and $q$. We do not include terms for $b$ quark mass mistuning since
the tuning is very accurate and we expect any small mistuning to have
negligible effect on bag parameters.

\subsection{Finite-volume, Strong Isospin-breaking and QED Effects}
\label{subsec:fvolsibqed}

Finite-volume effects can be estimated based on chiral perturbation theory.
The results in~\cite{Arndt:2004bg} show finite-volume effects
of $\mathcal{O}(1\%)$ for the bag parameters of $O_1$ in small lattice
volumes of size $L=$ 2.5 fm for physical $u/d$ quark masses.
On our much larger lattices, with a minimum size of $L=$ 4.6 fm at
physical $u/d$ masses for set 3, finite-volume effects will be a lot
smaller.  We conclude that this is a negligible effect at our current
level of uncertainties.

Strong-isospin breaking and electromagnetic effects can also be estimated
to be negligible at present. Our bag parameters show very little sensitivity
to the $u/d$ quark mass and our ratios for $B_s$ to $B_d$ differ from
1 by at most 10\% (Table~\ref{tab:bagparam}).
This suggests that changing $m_l$ to $m_d$ should only
have a $\mathcal{O}(0.1\%)$ effect.
Effects from the fact that the valence quarks have electromagnetic charge
are estimated at below 0.1\% for the decay constants in~\cite{Bazavov:2017lyh}.
They come largely from QED effects on the tuning of quark masses.
Since bag parameters are less sensitive to both heavy and light quark
masses than decay constants, we conclude that QED effects on the bag parameters
will be less than 0.1\% and we neglect them. Note that QED effects can still
enter $\Delta M_q$ or $\text{Br}(B_q \rightarrow \mu^+\mu^-)$ through
corrections to these processes from adding photons; these effects
need to be considered separately.

%%%%%%%%%%%%%%%%%%%%%%%%%%%%%%%%%%%%%%%%%%%%%%%%%%%%%%%%%%%%%%%%%%%%
\section{Correlations in Final Results}
\label{sec:corr}
In this Appendix we describe the correlations between the uncertainties
in different final results from our analysis.
Our principal results consist of values for
the reduced matrix elements of the 4-quark operators, $R^n_q$, defined
in Eq.~(\ref{eq:red-mat-element}),
evaluated for physical quark masses (Table~\ref{tab:OnMSB}). The results
for a given meson and different operators
are only weakly correlated, as shown in Table~\ref{tab:corr-os}
for the $B_s$~meson. There is more correlation, but still small,
between values of the bag parameters (\Eq{bagparam}),
because of the normalization factors~$\eta_n^s$ (\Eq{etadef}).
The means and standard deviations for these quantities are
collected in Table~\ref{tab:values}.

\begin{table}
\caption{\label{tab:corr-os} Correlations in the uncertainties of the
$R_{s}^{n}$ (\Eq{red-mat-element}) for different values of~$n$.
Correlations are also shown
for the bag parameters $B_{B_s}^{(n)}$ (\Eq{bagparam}),
and for the ratio $R^n_s/R^n_d$. Correlations
for $B_{B_s}^{(n)}/B_{B_d}^{(n)}$ are the same as
for $R^n_s/R^n_d$.
}

\begin{ruledtabular}
\begin{tabular}{crrrrr}
&$R_s^1$ & $R_s^2$ & $R_s^3$ & $R_s^4$ & $R_s^5$  \\
\hline\hline
$R_s^1$& $1.000$ & $-0.069$ & $0.013$ & $0.041$ & $0.040$  \\
$R_s^2$&& $1.000$ & $-0.039$ & $-0.040$ & $-0.026$  \\
$R_s^3$&&& $1.000$ & $0.023$ & $0.012$  \\
$R_s^4$&&&& $1.000$ & $0.144$  \\
$R_s^5$&&&&& $1.000$  \\
\hline\hline\\[-1ex]
&$B_{B_s}^{(1)}$ & $B_{B_s}^{(2)}$ & $B_{B_s}^{(3)}$ & $B_{B_s}^{(4)}$ & $B_{B_s}^{(5)}$  \\
\hline\hline
$B_{B_s}^{(1)}$& $1.000$ & $0.062$ & $0.012$ & $0.037$ & $0.038$  \\
$B_{B_s}^{(2)}$&& $1.000$ & $0.177$ & $0.233$ & $0.150$  \\
$B_{B_s}^{(3)}$&&& $1.000$ & $0.170$ & $0.107$  \\
$B_{B_s}^{(4)}$&&&& $1.000$ & $0.256$  \\
$B_{B_s}^{(5)}$&&&&& $1.000$  \\
\hline\hline\\[-1ex]
&$R_s^1/R_d^1$ & $R_s^2/R_d^2$ & $R_s^3/R_d^3$ & $R_s^4/R_d^4$ & $R_s^5/R_d^5$  \\
\hline\hline
$R_s^1/R_d^1$& $1.000$ & $0.296$ & $-0.034$ & $0.064$ & $0.047$  \\
$R_s^2/R_d^2$&& $1.000$ & $0.144$ & $0.068$ & $0.045$  \\
$R_s^3/R_d^3$&&& $1.000$ & $0.035$ & $0.018$  \\
$R_s^4/R_d^4$&&&& $1.000$ & $0.350$  \\
$R_s^5/R_d^5$&&&&& $1.000$  \\
\end{tabular}
\end{ruledtabular}

\end{table}

\begin{table}
\caption{\label{tab:values} Means and standard deviations for the reduced matrix elements $R^n_s$
and bag parameters $B^{(n)}_s$ for each 4-quark operator $O_n$, together with
values for their $B_s/B_d$~ratios.}

\begin{ruledtabular}
\begin{tabular}{ccccc}
& $R^n_s$& $R^n_s/R^n_d$& $B^{(n)}_s$& $B^{(n)}_s/B^{(n)}_d$ \\ \hline\hline
$O_1$ & $2.1678(928)$& $1.0081(250)$& $0.8129(348)$& $1.0081(250)$ \\
$O_2$ & $-2.1801(1035)$& $1.0589(242)$& $0.8169(431)$& $1.0626(243)$ \\
$O_3$ & $0.4357(288)$& $1.0886(339)$& $0.8163(572)$& $1.0924(340)$ \\
$O_4$ & $3.6532(1480)$& $0.9558(213)$& $1.0332(471)$& $0.9589(214)$ \\
$O_5$ & $1.9448(759)$& $0.9650(232)$& $0.9406(384)$& $0.9668(233)$ \\
\end{tabular}
\end{ruledtabular}

\end{table}

Values of $R^n_s$ are highly correlated with values
of $R^n_d$, for the same~$n$, which is why the ratios
${R_s^{n}}/{R_d^{n}}$
% \begin{equation}\label{eq:ratios}
%     r^{n} \equiv \frac{R_s^{n}}{R_d^{n}}
% \end{equation}
have much smaller uncertainties. Uncertainties in these ratios are
almost uncorrelated, however, with those in
the $R^n_q$ (correlations are~0.06 or smaller). Thus correlations
for the $B_d$~matrix elements~$R^n_d$ can be easily constructed from
the results in Table~\ref{tab:values} and
Table~\ref{tab:corr-os} for~$R^n_s$ and~$R^n_s/R^n_d$.
The ratio of bag parameters $B_{B_s}^{(n)}/B_{B_d}^{(n)}$ is
almost equal to~$R^n_s/R^n_d$ (the difference being only from small quark
mass effects) and has the same correlation matrix.

%%%%%%%%%%%%%%%%%%%%%%%%%%%%%%%%%%%%%%%%%%%%%%%%%%%%%%%%%%%%%%%%%
 %%%%%%%%%%%%%%%%%%%%%%%%%%%%%%%%%%%%%%%%%%%%%%%%%%%%%%
\section{SVD Cuts}\label{sec:SVD-cuts}
\subsection{The Problem}
There are three inputs for our least-square fits to sets
of correlators: 1) a collection of
$N_s$~random (Monte Carlo) samples $\Gv^{(s)}$, where each sample is
packaged as an $N_G$-dimensional vector; 2) a (vector)
fitting function
$\Gv(\pv)$ of fit parameters~$\pv$; and 3) \textit{a priori} estimates
(priors) for the fit parameters.

In the fit, the sample average
\begin{equation}
    \overline G \equiv \frac{1}{N_s} \sum_s \Gv^{(s)}
\end{equation}
is assumed to be a random sample drawn from a Gaussian distribution
with mean $\Gv(\pv^*)$ for some set $\pv^*$ of fit parameters, and a
covariance matrix given approximately by
\begin{align}
    \label{eq:cov}
    \cov &\approx \frac{1}{N_s(N_s-1)} \sum_s (\Gv^{(s)} - \overline \Gv)
        (\Gv^{(s)} - \overline \Gv)^T \\
        & \equiv \D \corr \D.
\end{align}
Here $\D$ is the diagonal matrix of standard deviations,
$\D_{ij} = \delta_{ij} \,\sigma_{\Gv_i}$, and
$\corr$ is the correlation matrix.
The best-fit parameters are obtained by minimizing
\begin{equation}
\label{eq:chi2}
    \chi^2(\pv) \equiv \sum_{n=1}^{N_G}
    \frac{\big((\overline\Gv - \Gv(\pv))^T \D^{-1} \vv_n\big)^2}{\lambda_n}
    + \chi^2_\mathrm{prior}
\end{equation}
as a function of the parameters~$\pv$, where $\lambda_n$ and $\vv_n$
are the eigenvalues and eigenvectors of the correlation matrix:
\begin{equation}
    \corr \vv_n = \lambda_n \,\vv_n .
\end{equation}
Note that
\begin{equation}
    \cov^{-1} = \sum_{n=1}^{N_G}
    \frac{\D^{-1} \vv_n \vv^T_n \D^{-1}}{\lambda_n}.
\end{equation}
$\chi^2_\mathrm{prior}(\pv)$ is the part of~$\chi^2(\pv)$ associated with the
Bayesian priors used in the fit.

The approximation for the covariance matrix, \Eq{cov},
 causes problems
if the number of samples~$N_s$ is insufficiently large
compared with the number of data
points~$N_G$~\cite{Michael:1993yj,Michael:1994sz}. In particular, the smaller
eigenvalues of the correlation matrix are underestimated. Indeed
it is obvious from \Eq{cov} that there must be $N_G-N_s$~modes
with zero eigenvalue when $N_s<N_G$. Underestimating eigenvalues
exaggerates their importance in
$\chi^2(\pv)$ (\Eq{chi2}), compromising the fit; and
$\chi^2(\pv)$ is undefined if there are zero eigenvalues.

\begin{figure}
\includegraphics[width=\hsize]{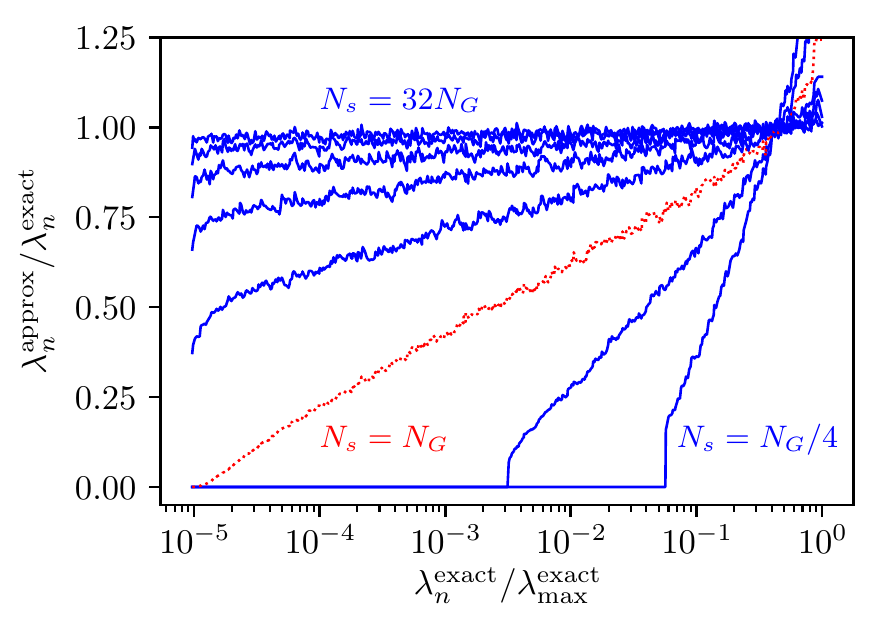}
\caption{\label{fig:ns-dependence} The ratio of approximate
to exact eigenvalues of the correlation matrix for $N_G=512$
correlated data points is plotted
versus the size of the exact eigenvalues divided by the maximum eigenvalue.
The approximate eigenvalues are determined (\Eq{cov})
from random samples of different
sizes ranging (by powers of~2) from $N_s=N_G/4$ to $N_s=32N_G$. The
red (dashed) line corresponds to $N_s=N_G$. Both sets of eigenvalues
(approximate and exact) are ordered from smallest to largest.
 }
\end{figure}

The underestimation of small eigenvalues
is illustrated in Figure~\ref{fig:ns-dependence},
which shows the ratio of
$\lambda_n^\mathrm{approx}/\lambda_n^\mathrm{exact}$ for approximate
eigenvalues estimated from random samples of different sizes
drawn from a simulation of a {known} distribution (so we know
the exact eigenvalues). The small eigenvalues are dragged down to zero by the
need for zero modes when $N_s<N_G$. They then increase slowly as new
samples are added, until
$\lambda_n^\mathrm{approx}/\lambda_n^\mathrm{exact}\approx1$ for all~$n$
when $N_s\gg N_G$. Note that good approximations
for all eigenvalues require~$N_s$ to
be 10--100~times
larger than~$N_G$. The figure shows results for $N_G=512$ pieces of correlated
data; curves for $N_G=50$ (or~$5000$) would be similar, but with more (or less)
noise. The range of values covered by the eigenvalues also has little effect
on the overall picture.

\subsection{Choosing an SVD Cut}

\begin{figure}
    \includegraphics[width=\hsize]{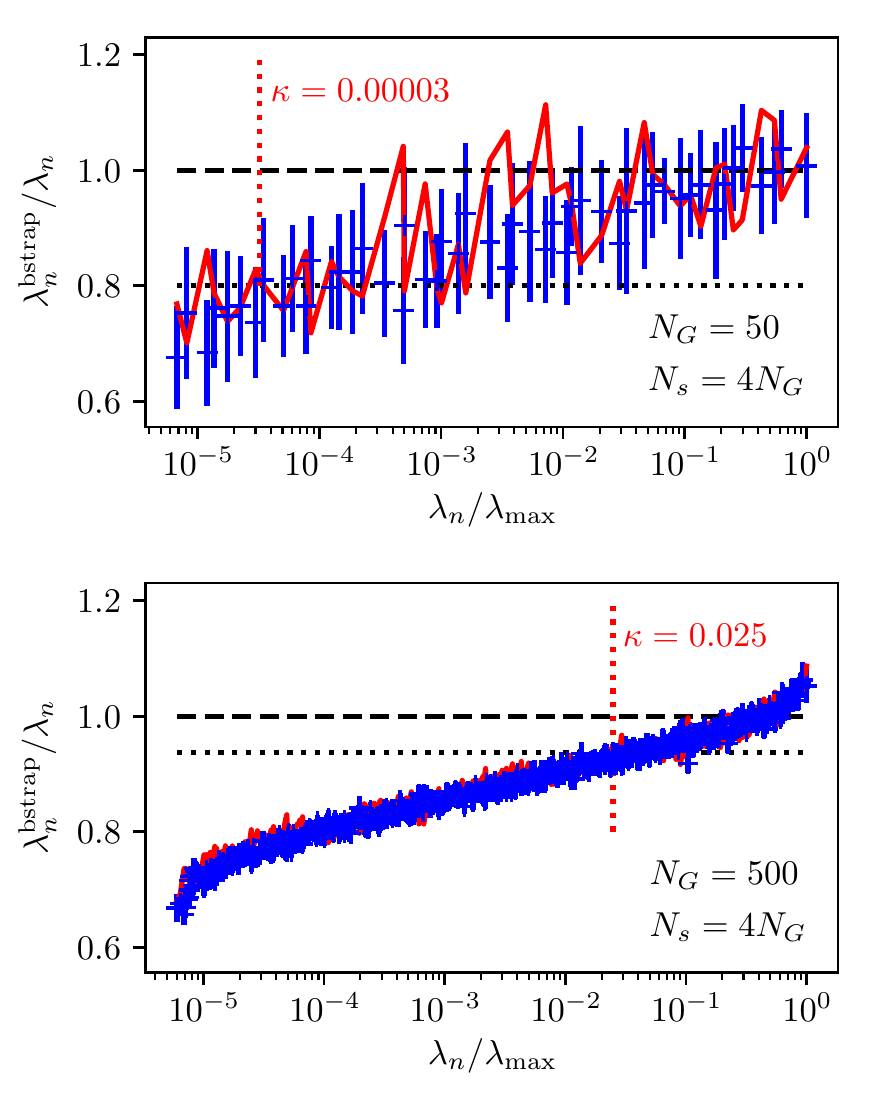}
    \caption{\label{fig:svd-bootstrap}
    Correlation-matrix eigenvalues computed from $N_s$ random samples
    of $N_G$ correlated data points are compared
    with eigenvalues computed from
    bootstrapped copies of the random sample. Results are shown
    for $N_G=50$ (top) and $N_G=500$ (bottom), with $N_s=4N_G$ in
    each case. The blue data points
    are ratios of bootstrapped eigenvalues to
    eigenvalues from the random sample itself; the error bars show the
    spread across different bootstrapped copies. The solid red
    line (mostly hidden in the bottom panel)
    shows ratios of eigenvalues from the random sample to
    those from the underlying distribution used to generate the random sample.
    The locations of the SVD cuts~$\kappa$ are shown by the vertical dashed
    red lines.}
\end{figure}

The problematic eigenvalues are those for which
\begin{equation}
\label{eq:svd-criterion}
    \frac{\lambda_n^\mathrm{approx}}{\lambda_n^\mathrm{exact}}
    < 1 - \sqrt{2/N_G}
\end{equation}
since, on average, individual terms in $\chi^2(\pv)$
should contribute approximately $1\pm\sqrt{2/N_G}$ to the total
(based on the width of the $\chi^2$~distribution).
Following~\cite{Michael:1993yj,Michael:1994sz}, we
deal with these eigenvalues by introducing a cutoff~$\kappa$
such that eigenvalues smaller than
$\kappa\lambda_\mathrm{max}$ are replaced with $\kappa\lambda_\mathrm{max}$, where
$\lambda_\mathrm{max}$ is the largest eigenvalue:
\begin{equation}
    \lambda_n \to \mathrm{max}(\lambda_n, \kappa\lambda_\mathrm{max}).
\end{equation}
Tuning~$\kappa$ appropriately,
this replacement increases the underestimated eigenvalues to a value that is
at least as large as the exact eigenvalue (and probably a lot larger).
Unlike in~\cite{Michael:1993yj,Michael:1994sz}, we do not renormalize
the eigenvalues to preserve the trace of the modified matrix (see below).

We need curves such as those in Figure~\ref{fig:ns-dependence} to set~$\kappa$,
but we don't know the exact eigenvalues in real applications.
An approximate curve can be generated
by comparing the eigenvalues of bootstrapped copies of our simulation
results~$\big\{\Gv^{(s)}\big\}$ with the eigenvalues from the full
ensemble. Each bootstrapped copy has $N_s$ samples, like the original
ensemble. In this analysis,
 bootstrapped eigenvalues play the
role of the approximate eigenvalues above, while eigenvalues computed
directly from  ensemble $\big\{\Gv^{(s)}\big\}$ now play the role of the
exact eigenvalues (since they specify the underlying distribution for
the bootstrapped copies).

Ratios of these eigenvalues are plotted in Figure~\ref{fig:svd-bootstrap}
(blue points) for examples with $N_G=50$~(top panel)
and $N_G=500$~(bottom panel) data points,
and $N_s=4N_G$ random samples for each data point. The error bars show
the spread of values across the different bootstrapped copies. These
points give us an approximate curve for $\lambda_n/\lambda_n^\mathrm{exact}$,
from which we can determine an SVD cutoff.

The ensembles used in these examples were generated from a known
distribution, so in this case we know the correct curve for
$\lambda_n^\mathrm{approx}/\lambda_n^\mathrm{exact}$\,---\,that is, the ratio
of the eigenvalues from the original ensemble to the exact eigenvalues
from the underlying distribution. The (solid) red line in the plots
shows this curve; it agrees well with the bootstrap estimates.

The vertical (dotted) red lines in each figure show the position
where the ratio curves intersect with $1-\sqrt{2/N_G}$ (bottom dotted line,
see \Eq{svd-criterion}).
We set the SVD cutoff at this point in each case. Fitting these data
we find that
\begin{equation}
    \chi^2/N_G =
    \begin{cases}
    0.90  & \mbox{with no SVD cut} \\
    0.82  & \mbox{with $\kappa=0.00003$}
    \end{cases}
\end{equation}
for $N_G=50$, showing that the SVD cut has a minimal effect (as expected), while
for $N_G=500$ we have
\begin{equation}
\label{eq:500-case}
    \chi^2/N_G =
    \begin{cases}
    1.30  & \mbox{with no SVD cut} \\
    0.41  & \mbox{with $\kappa=0.025$},
    \end{cases}
\end{equation}
which shows that the SVD cut is essential since $\chi^2/N_G=1.30$
is much too large for $N_G=500$\,---\,it corresponds to a $p$-value
of order~$3\times10^{-5}$.

\subsection{Conceptual Framework}
The nature of the SVD modification can be understood by representing
the ensemble-averaged data as a vector of Gaussian random variables,
\begin{equation}\label{eq:G-data}
    \Gv = \overline\Gv + \delta\Gv,
\end{equation}
where
\begin{equation}
    \delta\Gv \equiv \sum_{n=1}^{N_G} z_n \sqrt{\lambda_n} \,\D \vv_n,
\end{equation}
and the uncorrelated random variables $z_n$ satisfy:
\begin{equation}
    \langle z_n \rangle =0 \quad\quad \langle z_n z_m \rangle = \delta_{nm}.
\end{equation}
Here $\delta \Gv$ represents the uncertainty associated with the ensemble
average:
\begin{equation}
    \langle\delta\Gv \delta\Gv^T \rangle
    = \sum_n \lambda_n \D \vv_n \vv_n^T \D = M_\mathrm{cov}.
\end{equation}

The effect of the SVD cut is to add more
uncertainty,~$\delta \Gv^\mathrm{SVD}$:
\begin{equation}
    \Gv \to \overline\Gv + \delta\Gv + \delta \Gv_\mathrm{SVD},
\end{equation}
where
\begin{equation}
    \delta\Gv_\mathrm{SVD} \equiv \sum_{\lambda_n < \kappa\lambda_\mathrm{max}}
    \tilde{z}_n \sqrt{\kappa\lambda_\mathrm{max} - \lambda_n}\, \D \vv_n,
\end{equation}
and $\tilde{z}_n$ are new random variables with zero mean and unit
covariance matrix. Then
\begin{align}
    \big\langle (\delta\Gv &+ \delta\Gv_\mathrm{SVD})(\delta\Gv + \delta\Gv_\mathrm{SVD})^T \big\rangle
    \nonumber \\
    &= \langle \delta\Gv\delta\Gv^T \rangle + \langle \delta\Gv_\mathrm{SVD}\delta\Gv_\mathrm{SVD}^T \rangle
    \nonumber \\
    &= \sum_n \mathrm{max}(\lambda_n,\kappa\lambda_\mathrm{max})\, \D \vv_n \vv_n^T \D
\end{align}
is the SVD-modified covariance data. The SVD noise discussed above is
a random sample drawn from the distribution described
by~$\delta\Gv_\mathrm{SVD}$.

These formulas underscore the fact that introducing an SVD cut
is a conservative
move: it always increases the uncertainties in the data. This would
not necessarily be the case if we renormalized the eigenvalues after
introducing the SVD cut, as is done in~\cite{Michael:1993yj,Michael:1994sz}.
In practice, however, the difference between the two approaches is small.

Finally note that another option in an SVD analyses is to discard modes
below the cutoff. This corresponds to setting $\lambda_n=\infty$
for these modes,
which is much larger than is reasonable, much too conservative.
We find that fits are more accurate
and more stable using the prescription outlined above.

\subsection{Goodness of Fit}\label{goodness-of-fit}
Note that $\chi^2/N_G=0.41$ in \Eq{500-case} is much smaller than expected for
$N_G=500$: one expects~$1.00(6)$ instead. The small value arises because
random fluctuations in~$\overline\Gv$ are characteristic of the
uncertainties in $\delta\Gv$, but not those in~$\delta\Gv_\mathrm{SVD}$.
We can demonstrate this by adding a random sample
to~$\overline\Gv$ drawn from the distribution
specified by~$\delta\Gv_\mathrm{SVD}$,
\begin{equation}
    \delta\Gv_\mathrm{SVD} \to \mathrm{sample}(\delta\Gv_\mathrm{SVD}) + \delta\Gv_\mathrm{SVD},
\end{equation}
and refitting. In the case of \Eq{500-case}, a typical fit with SVD noise
gives $\chi^2/N_G$ increases to~0.96,
which is consistent with expectations.

Parenthetically, we note that overly broad priors\,---\,for example,
$0\pm10$ for a set of parameters that are all order~1\,---\,can also
result in a  small~$\chi^2$. This situation is addressed in a
similar fashion, by replacing the prior distribution~$\Pv$:
\begin{align}
    \Pv &\equiv \overline\Pv + \delta\Pv \nonumber \\
    &\to \overline\Pv + \mathrm{sample}(\delta\Pv) + \delta\Pv.
\end{align}
A good fit should have $\chi^2/N_G \approx 1\pm\sqrt{2/N_G}$ when
both SVD and prior
noise is included, and the fit results should agree (within errors) with
the results without noise.

A more direct test of a fitting protocol than adding extra noise
is to replace the fit data (\Eq{G-data}) with simulated data,
\begin{equation}
    \Gv_\mathrm{sim} \equiv
    \Gv(\pv_\mathrm{sim}) + \mathrm{sample}(\delta \Gv) + \delta\Gv,
\end{equation}
which has the same covariance matrix (from $\delta\Gv$) as
the real data, but whose mean is a random sample drawn from a distribution
whose mean is known ($\Gv(\pv_\mathrm{sim})$). A good fit to this simulated
data should give best-fit results for the parameters that agree with
$\pv_\mathrm{sim}$ to within errors. An obvious choice for the
simulation parameters~$\pv_\mathrm{sim}$ are the best-fit results obtained
when fitting the real data.

\bibliography{mix}

\end{document}